\documentclass[12pt,draftclsnofoot,onecolumn]{IEEEtran}
\usepackage{cite}
\usepackage{amsmath,amssymb,amsfonts}
\usepackage{algorithm}
\usepackage{algorithmic}
\usepackage{CJK}
\usepackage{longtable,booktabs}
\usepackage{booktabs}
\usepackage{graphicx}
\usepackage{subfigure}
\usepackage{textcomp}
\usepackage{flushend}
\usepackage{amsthm}
\usepackage{upgreek}
\usepackage{graphicx}
\usepackage{supertabular}
\usepackage[bottom]{footmisc}
\usepackage{stfloats}
\usepackage{color}
\usepackage{booktabs}
\usepackage{multirow}
\usepackage{array}
\usepackage{color}
\usepackage{epstopdf}
\UseRawInputEncoding 

\def\BibTeX{{\rm B\kern-.05em{\sc i\kern-.025em b}\kern-.08em
    T\kern-.1667em\lower.7ex\hbox{E}\kern-.125emX}}

\newcommand\bfb{\ensuremath{{\mathbf b}}}
\newcommand\bfB{\ensuremath{{\mathbf B}}}
\newcommand\bfd{\ensuremath{{\mathbf d}}}
\newcommand\bfD{\ensuremath{{\mathbf D}}}
\newcommand\bfA{\ensuremath{{\mathbf A}}}
\newcommand\bfg{\ensuremath{{\mathbf g}}}

\newcommand\bfS{\ensuremath{{\mathbf S}}}

\newcommand\bff{\ensuremath{{\mathbf f}}}
\newcommand\bfH{\ensuremath{{\mathbf H}}}
\newcommand\bfy{\ensuremath{{\mathbf y}}}
\newcommand\bfr{\ensuremath{{\mathbf r}}}
\newcommand\bfR{\ensuremath{{\mathbf R}}}
\newcommand\bfY{\ensuremath{{\mathbf Y}}}
\newcommand\bfx{\ensuremath{{\mathbf x}}}
\newcommand\bfX{\ensuremath{{\mathbf X}}}
\newcommand\bfz{\ensuremath{{\mathbf z}}}

\newcommand\bfv{\ensuremath{{\mathbf v}}}
\newcommand\bfw{\ensuremath{{\mathbf w}}}
\newcommand\bfW{\ensuremath{{\mathbf W}}}
\newcommand\bfV{\ensuremath{{\mathbf V}}}
\newcommand\bfa{\ensuremath{{\mathbf a}}}
\newcommand\bfn{\ensuremath{{\mathbf n}}}
\newcommand\bfN{\ensuremath{{\mathbf N}}}
\newcommand\bfI{\ensuremath{{\mathbf I}}}

\newcommand\bfQ{\ensuremath{{\mathbf Q}}}
\newcommand\bfF{\ensuremath{{\mathbf F}}}
\newcommand\rmT{\ensuremath{{\mathrm T}}}

\newcommand\rmU{\ensuremath{{\mathrm U}}}
\newcommand\rmD{\ensuremath{{\mathrm D}}}

\newcommand\rmH{\ensuremath{{\mathrm H}}}
\newcommand\rmIB{\ensuremath{{\mathrm {IB}}}}
\newcommand\rmLoS{\ensuremath{{\mathrm {LoS}}}}
\newcommand\rmNLoS{\ensuremath{{\mathrm {NLoS}}}}

\newcommand\rmB{\ensuremath{{\mathrm {B}}}}
\newcommand\rmBS{\ensuremath{{\mathrm {BS}}}}
\newcommand\rmdiag{\ensuremath{{\mathrm {diag}}}}

\newcommand\rmS{\ensuremath{{\mathrm S}}}
\newcommand\rmI{\ensuremath{{\mathrm I}}}
\newcommand\rmh{\ensuremath{{\mathrm h}}}

\newcommand\rmtr{\ensuremath{{\mathrm {tr} }}}

\newcommand\calO{\ensuremath{{\mathcal O}}}

\newcommand\calN{\ensuremath{{\mathcal N}}}

\newcommand\calM{\ensuremath{{\mathcal M}}}
\newcommand\calA{\ensuremath{{\mathcal A}}}

\newcommand\calL{\ensuremath{{\mathcal L}}}
\newcommand\bbC{\ensuremath{{\mathbb C}}}
\newcommand\bbR{\ensuremath{{\mathbb R}}}
\newcommand\bdzeta{\ensuremath{{\boldsymbol\zeta}}}
\newcommand\bdgamma{\ensuremath{{\boldsymbol\gamma}}}
\newcommand\bdtheta{\ensuremath{{\boldsymbol\theta}}}
\newcommand\bdomega{\ensuremath{{\boldsymbol\omega}}}

\begin{document}

\title{{Extreme Learning Machine}-based \\Channel Estimation in IRS-Assisted \\Multi-User ISAC System}

\author{\IEEEauthorblockN{
Yu Liu,
Ibrahim Al-Nahhal, \textit{Senior Member}, \textit{IEEE}, \\
Octavia A. Dobre, \textit{Fellow}, \textit{IEEE},
Fanggang Wang, \textit{Senior Member}, \textit{IEEE}, \\
\textcolor{black}{and Hyundong Shin, \textit{Fellow}, \textit{IEEE}}
}
\thanks{


Yu Liu and Fanggang Wang are with the State Key Laboratory of Advanced Rail Autonomous Operation, Frontiers Science Center for Smart High-speed Railway System, Beijing Jiaotong University, Beijing 100044, China (e-mail: yuliu1@bjtu.edu.cn; wangfg@bjtu.edu.cn).

Ibrahim Al-Nahhal and Octavia A. Dobre are with the Faculty of Engineering and Applied Science, Memorial University, St. John’s, NL A1C 5S7, Canada (e-mail: ioalnahhal@mun.ca; odobre@mun.ca).

Hyundong Shin is with the Department of Electronics and Information Convergence Engineering, Kyung Hee University, Yongin 17104, Republic of Korea (e-mail: hshin@khu.ac.kr).

{Digital Object Identifier: 10.1109/TCOMM.2023.3308150}

{This article is available at: https://ieeexplore.ieee.org/document/10229204}
}
}

\maketitle

\begin{abstract}
  Multi-user integrated sensing and communication (ISAC) assisted by intelligent reflecting surface (IRS) has been recently investigated to provide a high spectral and energy efficiency transmission.
  This paper proposes a practical channel estimation approach for the first time to an IRS-assisted multi-user ISAC system.
  The estimation problem in such a system is challenging since the sensing and communication (SAC) signals interfere with each other, and the passive IRS lacks signal processing ability.
  A two-stage approach is proposed to transfer the overall estimation problem into sub-ones, successively including the direct and reflected channels estimation.
  Based on this scheme, the ISAC base station (BS) estimates all the SAC channels associated with the target and uplink users, while each downlink user estimates the downlink communication channels individually.
  Considering a low-cost demand of the ISAC BS and downlink users, the proposed two-stage approach is realized by an efficient \textcolor{black}{neural network (NN)} framework that contains two different extreme learning machine (ELM) structures to estimate the above SAC channels.
  Moreover, two types of input-output pairs to train the ELMs are carefully devised, which impact the estimation accuracy and computational complexity under different system parameters.
  \textcolor{black}{Simulation results reveal \textcolor{black}{a} substantial \textcolor{black}{performance} improvement achieved by the proposed ELM-based approach over the least-squares and \textcolor{black}{NN}-based benchmarks, \textcolor{black}{with reduced training complexity and faster training speed}.}

\end{abstract}

\section{Introduction}\label{sec:intro}


\IEEEPARstart{T}{he} intelligent reflecting surface (IRS) technology has attracted significant research attention to boost the coverage and resource utilization efficiency of the next wireless system generations \cite{ref-new-IRS-defi,ref:IRS-SCMA-optimize,ref:IRS-SCMA}. 
\textcolor{black}{IRS consists of numerous passive electromagnetic reflecting elements \cite{ref-new-IRS-defi}.} 
Based on the surrounding wireless channel state information (CSI), each element independently alters its phase-shift to manipulate the transmission direction of the incoming signals.
By jointly configuring all the IRS elements, a desirable transmission environment is obtained, and the system performance can be further improved.
This technique is the so-called passive beamforming \cite{ref:ChModel-refpower,ref:new-IRS-passive-beam,ref:IRS-beam-gain3}. 
Since reliable beamforming requires accurate CSI, the channel estimation is essential in such IRS-assisted wireless systems.
Although the channel estimation methods for the conventional wireless systems have been widely investigated, they are difficult to be directly applied in the IRS-assisted systems due to two main challenges.
\textcolor{black}{First, unlike the traditional base station (BS) or user equipment (UE), the passive IRS lacks signal processing capacity.
In such a case, the channels of the IRS to BS/UEs are not available separately, and only the reflected channel (e.g., BS-IRS-UE link) can be jointly estimated.
Second, the reflected channel matrix has a large dimension and does not follow the Rayleigh distribution model.
This makes the optimal minimum mean square error channel estimator intractable to be implemented due to the multidimensional integration and extensive computational cost, while the available LS estimator reveals unsatisfactory estimation performance.
The above} two challenges lead to limited estimation accuracy and large training overhead demands for channel estimation in IRS-assisted wireless systems \cite{ref:IRS-turn-off,ref:new-IRS-Challenge}.  
{Recently, a variety of model-driven estimation approaches have been investigated to address the above challenges, including the binary reflection pattern controlled \cite{ref:IRS-ChE-onoff1}, discrete Fourier transform (DFT) protocol-based \cite{ref:new-IRS-ChE-DFT}, and element grouping \cite{ref:IRS-ChE-elegroup2} methods. 
\textcolor{black}{Despite the contributions of these works, the above challenges have not been
sufficiently addressed yet.
Hence, the data-driven \textcolor{black}{neural network (NN)} estimation approaches have been further researched in \cite{ref:DL-IRS-ChE-WCL,ref:IRS-ChE-CNN-group,ref:ELM-OFDM-CE,ref:DL-IRS-ChE-TWC}. 
Thanks to the powerful learning ability of the \textcolor{black}{NN}, such as convolutional neural network (CNN) \cite{ref:DL-IRS-ChE-WCL, ref:IRS-ChE-CNN-group}, extreme learning machine (ELM) \cite{ref:ELM-OFDM-CE}, and deep residual learning \cite{ref:DL-IRS-ChE-TWC}, {the} mapping between the received signals and channels can be successfully described to realize a satisfactory estimation accuracy and an acceptable training overhead.}


Integrated sensing and communication (ISAC) has been envisioned as another revolutionary technology to enhance the spectral utilization efficiency and reduce the hardware infrastructure cost for the next wireless system generations \cite{ref:ISAC-survey,ref:ChModel-b}. 
ISAC attempts to design a joint system that performs the sensing and communication (SAC) functionalities simultaneously.
The sensing functionality aims to detect the presence/absence of the targets and estimate their essential parameters (e.g., distance and angle) by using the interfered observations.
On the other hand, the communication functionality focuses on accurately recovering the transmitted information from the received noisy signals. 
\textcolor{black}{As such, an essential research topic is to pursue a trade-off between SAC functionalities and obtain their mutual performance gain in the ISAC systems \cite{ref:ISAC-improve-S1,ref:ISAC-improve-S2,ref:new-ISAC-improve-S2, ref:ISAC-improve-C1,ref:ISAC-improve-C2,ref:new-ISAC-improve-C3}.} 
\textcolor{black}{The communication-assisted} sensing was investigated in \cite{ref:ISAC-improve-S1,ref:ISAC-improve-S2, ref:new-ISAC-improve-S2}, which shed light on optimizing the sensing metrics, such as the target angle estimation accuracy and target detection probability, under the constraints of acceptable communication performance. 
On the contrary, the sensing-assisted communication is another research direction to enhance communication efficiency/reliability, including the sensing-assisted beamforming, tracking, and prediction \cite{ref:ISAC-improve-C1,ref:ISAC-improve-C2, ref:new-ISAC-improve-C3}.
The above literature generally assumed that perfect CSI is known at the receiver side while rarely \textcolor{black}{considered} the channel estimation problem for the ISAC systems.
To overcome this drawback, the authors in \cite{ref:ISAC-ChEst-RSU} demonstrated a sensing-assisted beamforming scheme, which jointly designed the beamforming of the roadside units (RSU) and vehicle.
With the aid of the radar-equipped RSU, the covariance matrices of both SAC channels required for the beamforming design were estimated by the echo signals.
However, the SAC channels in this scheme were assumed to share the same dominant paths.
Thus, the estimation approach in \cite{ref:ISAC-ChEst-RSU} is only limited to this particular case.

Motivated by the \textcolor{black}{profound} potential of both IRS and ISAC in enhancing the resource utilization efficiency, IRS is expected to support the ISAC in providing better SAC performance.
Recently, the interplay of SAC functionalities in the IRS-assisted ISAC systems was explored in \cite{ref:ChModel-pathloss-SJ,ref:Joint-ISAC-IRS-beam, ref:new-Joint-ISAC-IRS-beam1, ref:IRS-ISAC-wavebeam1,ref:IRS-ISAC-wavebeam2}.
The authors in \cite{ref:ChModel-pathloss-SJ,ref:Joint-ISAC-IRS-beam, ref:new-Joint-ISAC-IRS-beam1} jointly devised the beamforming for the ISAC BS and passive IRS to improve the sensing metric while ensuring \textcolor{black}{an} acceptable communication performance.
Moreover, the studies of waveform and passive beamforming co-design for the IRS-assisted ISAC systems were performed in \cite{ref:IRS-ISAC-wavebeam1} and \cite{ref:IRS-ISAC-wavebeam2}, considering the trade-off between SAC performance.
It should be noted that the above SAC performance improvement builds on {an} accurate CSI.
\textcolor{black}{Due to the mutual interference between the SAC signals, the channel estimation problem in such IRS-assisted ISAC systems is challenging.}
To overcome this problem, a CNN-based three-stage channel estimation approach was investigated in \cite{ref:IRS-ISAC-ChEst-TVT}, which estimated the sensing and uplink communication channels for an IRS-assisted single-user ISAC system.
This work was the first attempt to estimate the SAC channels in such a system;
however, it was limited to an uplink single-user scenario.
To the best of the authors' knowledge, the SAC channels estimation problems in multi-user IRS-assisted ISAC systems have not been addressed yet, and \textcolor{black}{efficient} {NN}-based estimation \textcolor{black}{solutions need to be developed.}

\textcolor{black}{To fulfill the research gap, this paper proposes a novel ELM-based \textcolor{black}{NN} channel estimation approach for an IRS-assisted multi-user ISAC system.
Different from the existing work in \cite{ref:IRS-ISAC-ChEst-TVT} that focused on the channel estimation of the target sensing and uplink single-user communication, this paper simultaneously estimates the channels of the sensed target, and uplink and downlink communication multi-users.
Correspondingly, their pilot transmission policies are also {different to accommodate the corresponding system model}.
Moreover, to build the \textcolor{black}{NN} estimation framework for the SAC channels, the ELM network is employed, rather than the CNN in \cite{ref:IRS-ISAC-ChEst-TVT}.
Note that the ELM is essentially a feedforward neural network (FNN) that involves a single hidden layer.}
\textcolor{black}{
Its distinctive advantages have been proved in dealing with various recognition/regression problems \cite{ref:ELM1,ref:EM-application1,ref:EM-application2}. 
First, the learning speed of the ELM is extremely fast, typically within a few seconds.
This lies in that the ELM network updates its parameters based on an inverse operation, without applying the gradient-based {backpropagation} learning algorithms.
Second, the learning and generalization capacities of the ELM are considerable, even when compared with the more complex \textcolor{black}{NNs}.
\textcolor{black}{The above advantages motivate the use of ELM in the considered SAC channels estimation problem.}}

\textcolor{black}{Motivated by the above facts, for the considered IRS-assisted multi-user ISAC system, a two-stage {channel} estimation approach is devised at both the full-duplex (FD) ISAC BS and downlink UEs.}
Then, an ELM-based \textcolor{black}{NN} framework is established correspondingly, along with input-output pairs designs.
The summary of the paper contributions is as follows:
\begin{enumerate}[]
  \item A two-stage estimation approach is proposed to estimate the SAC channels of the IRS-assisted multi-user ISAC system.
      By altering the IRS on/off state, it transfers the overall estimation problems into sub-ones, successively including the direct and reflected channels estimation.
      \textcolor{black}{In} this successive manner, the FD ISAC BS estimates all the sensing, UE-BS, and UE-IRS-BS channels, while each downlink UE estimates its BS-UE and BS-IRS-UE channels.
      The proposed approach enables a practical channel estimator for such a system for the first time and successfully addresses the estimation challenge caused by the inherent interference.
  \item 
      \textcolor{black}{The pilot transmission policy for the proposed two-stage approach is carefully devised to support the SAC channels estimation.
      In order to mitigate the effect of the inherent interference on the estimation performance and distinguish the SAC channels associated with the different equipment (i.e., ISAC BS, target, and multiple uplink/downlink UEs), the proposed pilot transmission policy involves the pilot signals design at the FD ISAC BS and uplink UEs, as well as the IRS phase-shift vectors design.}
  \item An ELM-based \textcolor{black}{NN} estimation framework is established at the ISAC BS and downlink UEs to estimate the SAC channels.
      Considering the system low-cost demands, two efficient ELM structures are designed to construct the proposed \textcolor{black}{NN} framework. 
      The first ELM is adopted to estimate the direct SAC channels in the first stage, while the second one is employed for the reflected channels estimation in the second stage.
  \item Two types of input-output pairs are proposed for the ELMs. 
      The first input-output pair is generated by the original received SAC signals, while the second one builds on the least-squares (LS) estimation results of SAC channels.
      The training samples are further enriched by leveraging data augmentation concepts to improve the estimation accuracy.
  \item The computational complexity of the inputs generation and ELM online testing for the proposed ELM-based approach is analyzed \textcolor{black}{in terms of number of real additions and multiplications}.
      Numerical results unveil that the proposed ELM approach achieves a comparable complexity to the LS benchmark estimator, satisfying the low-cost requirements of both ISAC BS and downlink UEs.
  \item Extensive simulations are conducted to verify the effectiveness of the proposed ELM-based approach under a wide range of signal-to-noise ratio (SNR) and system parameters setups.
      \textcolor{black}{Quantitative results demonstrate that the proposed ELM-based approach outperforms the LS and \textcolor{black}{NN}-based benchmark schemes in {terms of} estimation accuracy.
      Moreover, it exhibits fast learning speed and considerable generalization capacity under various SNR conditions.}
\end{enumerate}

The rest of the paper is organized as follows:
Section \ref{sec:system} presents the IRS-assisted multi-user ISAC system model.
Sections \ref{sec:ELM-approach} and \ref{sec:complexity} introduce the proposed ELM-based channel estimation approach and analyze its computational complexity, respectively.
Simulation results and conclusions are provided in Sections \ref{sec:simulation} and \ref{sec:conclusion}, respectively.

{\textit{Notations}}: Boldface lowercase and uppercase letters stand for vectors and matrices, respectively.
$\mathcal{CN}(\mu,\sigma^2)$ represents a circularly symmetric complex Gaussian random variable with mean $\mu$ and variance $\sigma^2$. 
$\bfI_M$ is an identity matrix of size $M$.
For any vector $\bfx$, $\text{diag}\{\bfx\}$ returns a diagonal square matrix whose diagonal consists of the elements of $\bfx$.
The operators $(\cdot)^{\rm H}$, $(\cdot)^\mathrm{T}$, $(\cdot)^{-1}$, $(\cdot)^\dag$, vec$[\cdot]$, $\Re\{\cdot\}$, $\Im\{\cdot\}$, $\mathbb E\{\cdot\}$, $\mathbb D\{\cdot\}$, $\|\cdot\|_2$, and $\|\cdot\|_F$ \textcolor{black}{correspond} to the Hermitian, transpose, inverse, pseudoinverse, vectorization, real part, imaginary part, expectation, variance, second order norm, and Frobenius norm of their arguments, respectively. 
$\jmath=\sqrt{-1}$ denotes the imaginary unit.
$\calN_a^b=\{a,a+1,\ldots,b\}$ {represents} the index set from integer $a$ to $b$, and $a<b$.


\begin{figure}
\centering
\subfigure[ ]
{\begin{minipage}[b]{0.37\textwidth}
\includegraphics[width=1\textwidth]{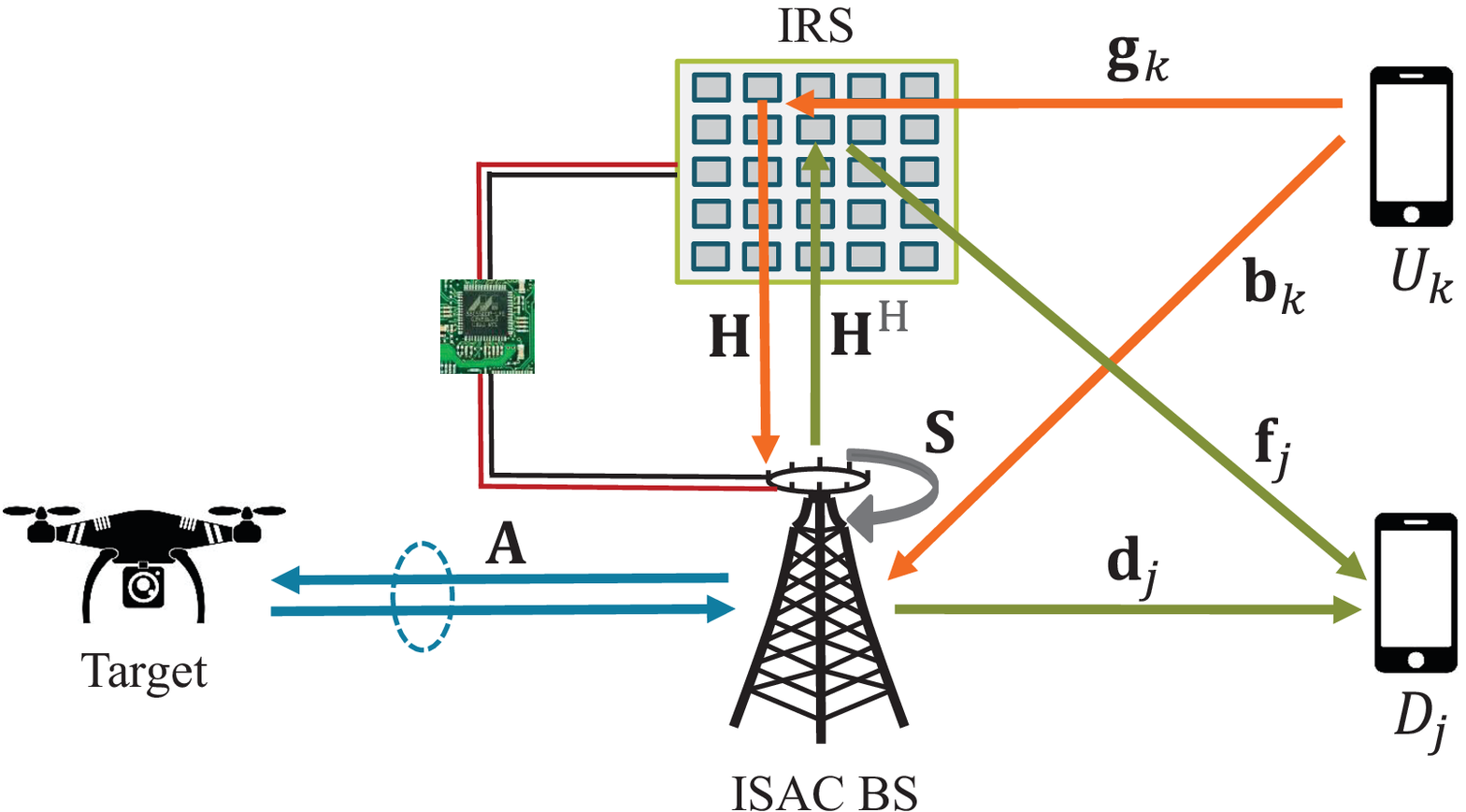} \quad
\end{minipage}}
\subfigure[ ]
{\begin{minipage}[b]{0.48\textwidth}
\includegraphics[width=1\textwidth]{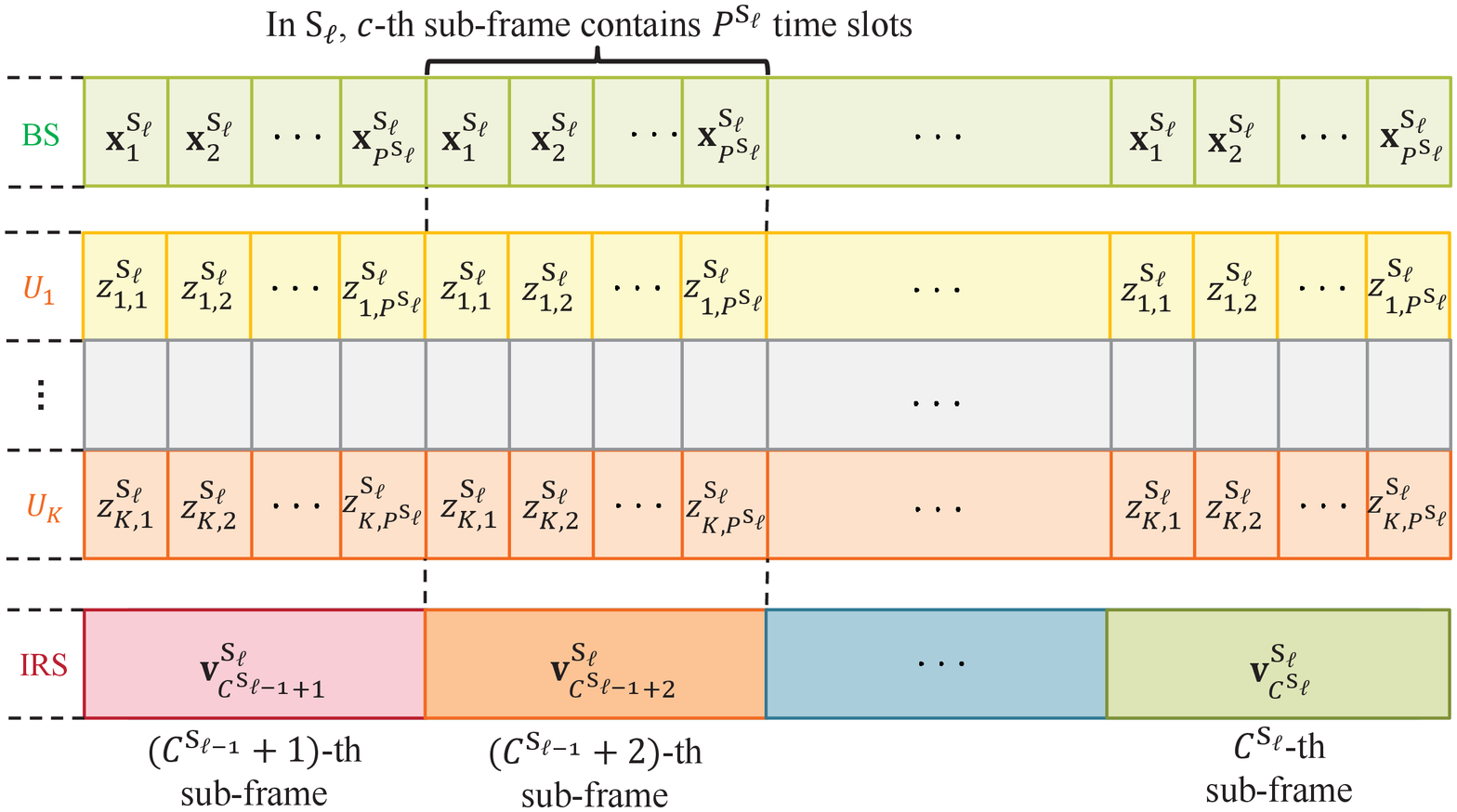}
\end{minipage}}
\caption{\textcolor{black}{IRS-assisted multi-user ISAC system: (a) System model, (b) Pilot transmission policy.}}
\label{fig:System}
\end{figure}

\section{System Model}\label{sec:system}
Consider a {narrowband} IRS-assisted multi-user ISAC system as illustrated in \textcolor{black}{Fig. \ref{fig:System}(a)}, which contains a FD ISAC BS, target, $K$ uplink UEs, $J$ downlink UEs, and an IRS.
Define $U_k$, $k\in\calN_1^K$, and $D_j$, $j\in\calN_1^J$, as the $k$-th uplink UE and $j$-th downlink UE, respectively.
The FD ISAC BS is equipped with $M$ transmit and $M$ receive antennas, while the UEs are \textcolor{black}{equipped} with single antenna.
For target sensing, the ISAC BS transmits the radar probing waveforms to the surrounding target and receives the echo signal from the BS-target-BS channel, $\bfA\in\bbC^{M\times M}$.
For communication, the IRS is equipped with $L$ passive reflecting elements, \textcolor{black}{which} is deployed to assist the ISAC BS in serving the UEs.
The uplink $U_k$ transmits the signals to the ISAC BS through the direct $U_k$-BS and reflected $U_k$-IRS-BS channels, while the downlink $D_j$ receives the signals from the ISAC BS through the direct BS-$D_j$ and reflected BS-IRS-$D_j$ channels.
Let $\bfb_k\in\bbC^{M\times 1}$, $\bfg_k\in\bbC^{L\times 1}$, $\bfH\in\bbC^{M\times L}$, $\bfd_j\in\bbC^{M\times 1}$, and $\bff_j\in\bbC^{L\times 1}$ denote the communication channel coefficients of the $U_k$-BS, $U_k$-IRS, IRS-BS, BS-$D_j$, and IRS-$D_j$ links, respectively.
\textcolor{black}{Note that all the SAC channels are assumed to be flat-fading considering the narrowband transmission \cite{ref:major-SI-estimation3,ref:major-SI-estimation4,ref:major-SI-estimation5}.} 
Since the ISAC BS operates in the FD mode, the self-interference (SI) from the SI channel, $\bfS\in\bbC^{M\times M}$, is induced to the ISAC BS.


\textcolor{black}{To estimate the SAC channels of the IRS-assisted multi-user ISAC system, a pilot transmission policy is devised and illustrated in \textcolor{black}{Fig. \ref{fig:System}(b)}.
It includes the pilot signals design at the FD ISAC BS and multiple uplink UEs, and the IRS phase-shift vectors design.
Let $\rmS_\ell$, $\ell\in\calN_1^2$, denote the $\ell$-th estimation stage in the proposed two-stage approach.
As shown in \textcolor{black}{Fig. \ref{fig:System}(b)}, $\rmS_\ell$ lasts from the $(C^{\rmS_{\ell-1}}+1)$-th to the $C^{\rmS_\ell}$-th sub-frames with $C^{\rmS_0}=0$.}
Then, in $\rmS_\ell$, the FD ISAC BS and uplink $U_k$, $k\in\calN_1^K$, simultaneously transmit the pilot signals for $C^{\rmS_\ell}-C^{\rmS_{\ell-1}}$ sub-frames.
The $c$-th, $c\in\calN_{C^{\rmS_{\ell-1}}+1}^{C^{\rmS_\ell}}$, sub-frame contains $P^{\rmS_\ell}$ time slots.
At the $p$-th, $p\in\calN_1^{P^{\rmS_\ell}}$, time slot in each sub-frame, let $\bfx_{p}^{\rmS_\ell}\in\bbC^{M\times 1}$ represent the pilot signal vector transmitted by the ISAC BS, while the pilot symbol adopted at uplink $U_k$ is denoted by $z_{k,p}^{\rmS_\ell}$.
Correspondingly, their pilot signal matrix and vector in one sub-frame \textcolor{black}{can be} written as $\bfX^{\rmS_\ell}=[\bfx_1^{\rmS_\ell},\bfx_2^{\rmS_\ell},\ldots,\bfx_{P^{\rmS_\ell}}^{\rmS_\ell}]\in\bbC^{M\times P^{\rmS_\ell}}$ and $\bfz_{k}^{\rmS_\ell}=[z_{k,1}^{\rmS_\ell},z_{k,2}^{\rmS_\ell},\ldots,z_{k,P^{\rmS_\ell}}^{\rmS_\ell}]\in\bbC^{1\times P^{\rmS_\ell}}$, respectively.
Moreover, the IRS phase-shift vector in the $c$-th sub-frame is defined as $\bfv_c^{\rmS_\ell}\in\bbC^{L\times 1}$, which keeps unchanged within one sub-frame.
\textcolor{black}{Without loss of generality, the pilot symbol duration (i.e., time slot duration), $T_{\rm P}$, is assumed to be much smaller than the channel coherence time, $T_{\rm coh}$, whereas it is larger than the propagation delays of the SAC and residual SI signals \cite{ref:major-3gpp.36.211}.
This assumption ensures that all the SAC channels are unchanged during the two estimation stages, while the received SAC and residual SI signals can be synchronized.
Based on the above pilot transmission policy, at the $p$-th time slot in the $c$-th sub-frame, the received signals at the ISAC BS and downlink $D_j$, $\bfy_{c,p}^{\rmS_\ell}\in\bbC^{M\times 1}$ and $r_{j,c,p}^{\rmS_\ell}$, are respectively expressed as}
\begin{align}
\bfy_{c,p}^{\rmS_\ell}
& = \underbrace{ \sum_{k\in\calN_1^K}\Big(\bfb_k + \bfH \rmdiag\{\bfv_c^{\rmS_\ell}\} \bfg_k \Big)z_{k,p}^{\rmS_\ell} }_{\text {Uplink communication signal}}  + \underbrace{ \bfA^\rmH\bfx_{p}^{\rmS_\ell} }_{\text {Sensing signal}} + \underbrace{ \bfS^\rmH\bfx_{p}^{\rmS_\ell} }_{\text {Residual SI}} + \bfn_{c,p}^{\rmS_\ell}, 
\label{eq:y_cp_Orig}
\end{align}
and 
\begin{align}
& r_{j,c,p}^{\rmS_\ell} = \underbrace{ \Big( \bfd_j^\rmH + \bff_j^\rmH \rmdiag\{(\bfv_c^{\rmS_\ell})^\rmH\} \bfH^\rmH \Big)\bfx_{p}^{\rmS_\ell} }_{\text {Downlink communication signal}} + w_{j,c,p}^{\rmS_\ell}, 
\label{eq:r_jcp_Orig}
\end{align}
where $\bfv_c^{\rmS_\ell}=[\beta_c e^{\jmath\phi_{c,1}}, \beta_c e^{\jmath\phi_{c,2}}, \ldots, $ $\beta_c e^{\jmath\phi_{c,L}}]^\rmT$ with $\beta_c\in[0,1]$ and $\phi_{c,l}\in[0,2\pi)$ with $l\in\calN_1^L$ are the amplitude and phase-shift of the $l$-th IRS element, respectively.
The additive white Gaussian noise $\bfn_{c,p}^{\rmS_\ell}\in\bbC^{M\times 1}$ and $w_{j,c,p}^{\rmS_\ell}$ respectively follow ${\cal CN}(0,\sigma^2\bfI_M)$ with zero-mean and variance \textcolor{black}{of} $\sigma^2$, and $\mathcal{CN}(0,\varsigma^2)$ with zero-mean and variance \textcolor{black}{of} $\varsigma^2$.
\textcolor{black}{
Moreover, assume that the SI channel $\bfS$ is slow-varying, and the residual SI follows the linear SI model in \cite{ref:major-3.5G,ref:ChModel-refpower}. 
Then, $\bfS$ can be pre-estimated at the ISAC BS, and the residual SI in \eqref{eq:y_cp_Orig} is {compensated} before estimating the SAC channels.
}
According to \eqref{eq:y_cp_Orig} and \eqref{eq:r_jcp_Orig}, it is obvious that $\bfH\rmdiag\{\bfv_c^{\rmS_\ell}\}\bfg_k=\bfH\rmdiag\{\bfg_k\}\bfv_c^{\rmS_\ell}$ and $\bfH\rmdiag\{\bfv_c^{\rmS_\ell}\}\bff_k=\bfH\rmdiag\{\bff_k\}\bfv_c^{\rmS_\ell}$.
The reflected channels of {the} $U_k$-IRS-BS and BS-IRS-$D_j$ links are equivalent to $\bfB_k = \bfH\rmdiag\{\bfg_k\}$ and $\bfD_j = \bfH\rmdiag\{\bff_j\}$, respectively.
As such, \eqref{eq:y_cp_Orig} and \eqref{eq:r_jcp_Orig} are respectively rewritten as
\begin{align}
& \bfy_{c,p}^{\rmS_\ell} =  \underbrace{ \sum_{k\in\calN_1^K}\Big( \bfb_k + \bfB_k\bfv_c^{\rmS_\ell} \Big)z_{k,p}^{\rmS_\ell} }_{\text {Uplink communication signal}} + \underbrace{ \bfA^\rmH\bfx_{p}^{\rmS_\ell} }_{\text {Sensing signal}} + \bfn_{c,p}^{\rmS_\ell}, 
\label{eq:y_cp}
\end{align}
and
\begin{align}
& r_{j,c,p}^{\rmS_\ell} = \underbrace{ \Big( \bfd_j^\rmH + (\bfv_c^{\rmS_\ell})^\rmH\bfD_j^\rmH \Big)\bfx_{p}^{\rmS_\ell} }_{\text {Downlink communication signal}} + w_{j,c,p}^{\rmS_\ell}. 
\label{eq:r_jcp}
\end{align}
Due to the mutual interference between SAC signals, the SAC channels estimation problems in \eqref{eq:y_cp} and \eqref{eq:r_jcp} are challenging.
In the following, based on the {proposed} pilot transmission policy in \textcolor{black}{Fig. \ref{fig:System}(b)}, we investigate \textcolor{black}{the estimation of} the SAC channels (i.e., $\bfA$, $\bfb_k$, and $\bfB_k$, $k\in\calN_1^K$) at the ISAC BS and downlink communication channels (i.e., $\bfd_j$ and $\bfD_j$, $j\in\calN_1^J$) at the downlink UEs for {the} IRS-assisted multi-user ISAC system.

\section{Proposed ELM-based Estimation Approach}\label{sec:ELM-approach}
This section proposes a novel \textcolor{black}{NN}-based SAC channels estimation approach.
A two-stage approach is firstly presented to simplify the estimation problem.
Then, the generation of the input-outputs pairs for the \textcolor{black}{NN} is designed for each estimation stage.
On this basis, an ELM-based \textcolor{black}{NN} estimation framework is further \textcolor{black}{utilized}.

\subsection{Two-stage Estimation Approach}
To decouple the channel estimation problems in \eqref{eq:y_cp} and \eqref{eq:r_jcp}, a two-stage estimation approach is designed here.
\textcolor{black}{By simply altering the on/off state of IRS, the direct SAC channels (i.e., $\bfA$, $\bfb_k$, and $\bfd_j$) are estimated in $\rmS_1$, while the reflected $U_k$-IRS-BS and BS-IRS-$D_j$ channels (i.e., $\bfB_k$ and $\bfD_j$) are estimated in $\rmS_2$.
It is worth mentioning that a smart IRS controller enables the ISAC BS to communicate with the IRS through the BS-IRS backhaul link.
In addition, it alters the IRS on/off state by controlling the biasing voltages and load resistance of the positive-intrinsic-negative diodes in the IRS elements \cite{ref:IRS-turn-off}.}

\subsubsection{First Stage}
To estimate the direct SAC channels (i.e., $\bfA$, $\bfb_k$, and $\bfd_j$), all the IRS elements are turned off during the $1$-st to the $C^{\rmS_1}$-th sub-frames and $C^{\rmS_1}\geq 1$. 
According to the pilot transmission policy in \textcolor{black}{Fig. \ref{fig:System}(b)}, the direct SAC signals in the $c$-th sub-frame, $\bfY_c^{\rmS_1}\in\bbC^{M\times P^{\rmS_1}}$ and $\bfr_{j,c}^{\rmS_1}\in\bbC^{1\times P^{\rmS_1}}$, are respectively received at the ISAC BS and downlink $D_j$ as
\begin{align}
\bfY_c^{\rmS_1} = \bfA^\rmH\bfX^{\rmS_1} + \sum_{k\in\calN_1^K} \bfb_k\bfz_{k}^{\rmS_1} + \bfN_c^{\rmS_1}, \quad c\in\calN_{1}^{C^{\rmS_1}},
\label{eq:Y_c}
\end{align}
and 
\begin{align}
\bfr_{j,c}^{\rmS_1} = \bfd_j^\rmH \bfX^{\rmS_1} + \bfw_{j,c}^{\rmS_1}, \quad j\in\calN_1^J, \quad c\in\calN_{1}^{C^{\rmS_1}},
\label{eq:r_jc}
\end{align}
where $\bfN_c^{\rmS_\ell}=[\bfn_{c,1}^{\rmS_\ell},\bfn_{c,2}^{\rmS_\ell},\ldots,\bfn_{c,P^{\rmS_\ell}}^{\rmS_\ell}]\in\bbC^{M\times P^{\rmS_\ell}}$ and $\bfw_{j,c}^{\rmS_\ell}=[w_{j,c,1}^{\rmS_\ell},w_{j,c,2}^{\rmS_\ell},\ldots,w_{j,c,P^{\rmS_\ell}}^{\rmS_\ell}]\in\bbC^{1\times P^{\rmS_\ell}}$ represent the noise matrix and vector in $\rmS_\ell$ \textcolor{black}{with $\ell=1$}, respectively.
To mitigate the mutual interference between SAC signals and distinguish different uplink UEs, the adopted pilot signals (i.e., $\bfX^{\rmS_1}$ and $\bfz_{k}^{\rmS_1}$, $k\in\calN_1^K$) should be orthogonal.
\textcolor{black}{Consider the} DFT matrix, $\bfQ^{\rmS_\ell}\in\bbC^{R^{\rmS_\ell}\times P^{\rmS_\ell}}$, \textcolor{black}{can be given} by
\begin{align}
\bfQ^{\rmS_\ell}=
\left[
\begin{array}{cccc}
1 & 1 & \cdots & 1 \\
1 & W_{P^{\rmS_\ell}}^1 & \cdots & W_{P^{\rmS_\ell}}^{P^{\rmS_\ell}-1} \\
\vdots & \vdots & \ddots & \vdots \\
1 & W_{P^{\rmS_\ell}}^{R^{\rmS_\ell}-1} & \cdots & W_{P^{\rmS_\ell}}^{(R^{\rmS_\ell}-1)(P^{\rmS_\ell}-1)}
\end{array}
\right],
\label{eq:Qc}
\end{align}
where $W_{P^{\rmS_\ell}}^{nu}=e^{\jmath\frac{2\pi}{P^{\rmS_\ell}}{(n-1)(u-1)}}$ is the $(n,u)$-th entry \textcolor{black}{of} $\bfQ^{\rmS_\ell}$. In $\rmS_1$, let $R^{\rmS_1}=M+K$ and $P^{\rmS_1}\geq M+K$ \cite{ref:Bayesian-pilot-num}.
Hence, $\bfX^{\rmS_1}$ can be generated by selecting the \textcolor{black}{$1$-st} to the $M$-th rows of $\bfQ^{\rmS_1}$ and multiplying $\frac{1}{\sqrt M}$ on them (i.e., transmit power normalization), whereas $\bfz_{k}^{\rmS_1}$ corresponds to the $(M+k)$-th row of $\bfQ^{\rmS_1}$.
Considering the requirement of low pilot overhead, let $P^{\rmS_1}=M+K$.

\subsubsection{Second Stage}
In $\rmS_2$, all the IRS elements are turned on during the $(C^{\rmS_1}+1)$-th to the $C^{\rmS_2}$-th sub-frames to estimate the reflected communication channels (i.e., $\bfB_k$ and $\bfD_j$).
Then, the SAC signals in the $c$-th sub-frame, $\bfY_c^{\rmS_2}\in\bbC^{M\times P^{\rmS_2}}$ and $\bfr_{j,c}^{\rmS_2}\in\bbC^{1\times P^{\rmS_2}}$, are respectively received at the ISAC BS and downlink $D_j$ as
\begin{align}
\bfY_c^{\rmS_2}
& = \sum_{k\in\calN_1^{K}} \Big(\bfb_k + \bfB_k\bfv_c^{\rmS_2}\Big)\bfz_{k}^{\rmS_2}  + \bfA^\rmH\bfX^{\rmS_2} + \bfN_c^{\rmS_2}, \quad c\in\calN_{C^{\rmS_1}+1}^{C^{\rmS_2}},
\label{eq:Y_c_S2}
\end{align}
and 
\begin{align}
\bfr_{j,c}^{\rmS_2}
& = \Big( \bfd_j^\rmH + {(\bfv_c^{\rmS_2})}^\rmH\bfD_j^\rmH\Big)\bfX^{\rmS_2}  + \bfw_{j,c}^{\rmS_2}, \quad j\in\calN_1^J, \quad c\in\calN_{C^{\rmS_1}+1}^{C^{\rmS_2}}.
\label{eq:r_jc_S2}
\end{align}
In $\rmS_2$, $\bfQ^{\rmS_2}$ in \eqref{eq:Qc} is designed as a DFT matrix with $R^{\rmS_2}=\max\{M, K\}$ and $P^{\rmS_2}\geq \max\{M, K\}$ \cite{ref:Bayesian-pilot-num}.
Consequently, the adopted $\bfX^{\rmS_2}$ is formed by utilizing the \textcolor{black}{$1$-st} to the $M$-th rows of $\bfQ^{\rmS_2}$ and multiplying $\frac{1}{\sqrt M}$ on them, while $\bfz_{k}^{\rmS_2}$ is equal to the $k$-th row of $\bfQ^{\rmS_2}$.
Considering the above design of $\bfX^{\rmS_2}$ and $\bfz_k^{\rmS_2}$, as well as the low pilot overhead demand, let $P^{\rmS_2}= \max\{M, K\}$.
Moreover, from the $(C^{\rmS_1}+1)$-th to the $C^{\rmS_2}$-th sub-frames, the IRS phase-shift matrix is defined as $\bfV^{\rmS_2}=[\bfv_{C^{\rmS_1}+1}^{\rmS_2},\bfv_{C^{\rmS_1}+2}^{\rmS_2},\ldots,\bfv_{C^{\rmS_2}}^{\rmS_2}] \in\bbC^{L\times (C^{\rmS_2}-C^{\rmS_1})}$.
\textcolor{black}{Here, $\bfV^{\rmS_2}$ is taken as a DFT matrix with $C^{\rmS_2}-C^{\rmS_1}\geq L$, as commonly adopted in the literature \cite{ref:new-IRS-ChE-DFT} and \cite{ref:DL-IRS-ChE-TWC}.}
This is regarded as an optimal solution to improve the received signal power at the ISAC BS and downlink UEs, and {to} ensure an accurate channel estimation \cite{ref:new-IRS-ChE-DFT}. 

\subsection{Input-Output Pairs Design} \label{sec:IO-design}
Based on the received SAC signals at the {FD} ISAC BS and downlink UEs, two types of input-output pairs are {proposed and} designed for the {proposed} ELMs in each estimation stage.
\textcolor{black}{Regarding the design concepts of the different types of inputs for the proposed ELM, the first one is built on the original received SAC signals, while the second one adopts the LS estimation results of the SAC channels.
The output design for the proposed ELM \textcolor{black}{approach} relies on the actual values of the SAC channels. }

\subsubsection{\textcolor{black}{First Stage-Direct SAC Channels Estimation}}
The direct SAC channels (i.e., $\bfA$ and $\bfb_k$) are estimated at the ISAC BS in $\rmS_1$.
\textcolor{black}{For a more detailed description, let $\rmI_t$ represent the $t$-th, $t\in\calN_1^2$, input-output pair type.}

\textcolor{black}{\textbf{First Type of Input:}} 
\textcolor{black}{The} first type of input for the proposed ELM \textcolor{black}{approach} at the ISAC BS, $\bdzeta_\rmBS^{\rmS_1\rmI_1}$, is constructed by the received signals in \eqref{eq:Y_c} as
\begin{align}
\bdzeta_\rmBS^{\rmS_1\rmI_1}
& =\Big[\Re \big\{{\rm{vec}} \big[\bfY_1^{\rmS_1},\ldots,\bfY_{C^{\rmS_1}}^{\rmS_1}\big] \big\},  \Im \big\{{\rm{vec}} \big[\bfY_1^{\rmS_1},\ldots,\bfY_{C^{\rmS_1}}^{\rmS_1}\big] \big\}\Big]^\rmT.
\label{eq:G_BS_S1I1}
\end{align}

\textcolor{black}{\textbf{Second Type of Input:}} The second type of input for the {proposed} ELM \textcolor{black}{approach} is built on the LS estimates of the direct SAC channels.
Note that $\bfX^{\rmS_1}$ and $\bfz_k^{\rmS_1}$ are designed as orthogonal pilot signals in $\rmS_1$.
Then, based on the model in \eqref{eq:Y_c}, the LS estimation results of the BS-target-BS and $U_k$-BS channels, $\bar\bfA$ and $\bar\bfb_k$, are respectively obtained at the ISAC BS as
\begin{align}
\bar\bfA = \mathbb E\big\{ ( \bfY_c^{\rmS_1}(\bfX^{\rmS_1})^\dag)^\rmH\big\} = \bfA + \mathbb E\big\{ (\bar\bfN_c^{\rmS_1})^\rmH\big\},
\label{eq:A_bar}
\end{align}
and
\begin{align}
\bar\bfb_k = \mathbb E\big\{ \bfY_c^{\rmS_1} (\bfz_k^{\rmS_1})^\dag\big\} = \bfb_k + \mathbb E\big\{ \tilde\bfN_c^{\rmS_1}\big\}, \quad k\in\calN_1^K,
\label{eq:b_bar}
\end{align}
where $(\bfX^{\rmS_1})^\dag=(\bfX^{\rmS_1})^\rmH (\bfX^{\rmS_1}(\bfX^{\rmS_1})^\rmH)^{-1}$, $(\bfz_k^{\rmS_1})^\dag=(\bfz_k^{\rmS_1})^\rmH (\bfz_k^{\rmS_1}(\bfz_k^{\rmS_1})^\rmH)^{-1}$, $\bar{\bfN}_c^{\rmS_1}={\bfN}_c(\bfX^{\rmS_1})^\dag$, and $\tilde{\bfN}_c^{\rmS_1}={\bfN}_c(\bfz_k^{\rmS_1})^\dag$.
Therefore, by using \eqref{eq:A_bar} and \eqref{eq:b_bar}, the second type of input for the {proposed} ELM \textcolor{black}{approach} at the ISAC BS, $\bdzeta_\rmBS^{\rmS_1\rmI_2}$, is denoted by
\begin{align}
\bdzeta_\rmBS^{\rmS_1\rmI_2}
& = \Big[\Re\big\{\big[{\rm{vec}}[ \bar\bfA], \bar\bfb_1^\rmT, \bar\bfb_2^\rmT, \ldots,\bar\bfb_K^\rmT \big]\big\},  \Im \big\{ \big[ {\rm{vec}}[\bar\bfA], \bar\bfb_1^\rmT, \bar\bfb_2^\rmT, \ldots,\bar\bfb_K^\rmT \big] \big\} \Big]^\rmT.
\end{align}

\textcolor{black}{\textbf{Output:}} Corresponding to the above two types of inputs (i.e, $\bdzeta_\rmBS^{\rmS_1\rmI_1}$ and $\bdzeta_\rmBS^{\rmS_1\rmI_2}$), the output of the ELM at the ISAC BS, $\bdgamma_\rmBS^{\rmS_1}$, is constructed by the actual values of the direct SAC channels (i.e., $\bfA$ and $\bfb_k$) as
\begin{align}
\bdgamma_\rmBS^{\rmS_1}
& = \Big[\Re\big\{\big[{\rm{vec}}[\bfA], \bfb_1^\rmT,\bfb_2^\rmT,\ldots,\bfb_K^\rmT \big]\big\},  \Im \big\{ \big[ {\rm{vec}}[\bfA], \bfb_1^\rmT,\bfb_2^\rmT,\ldots,\bfb_K^\rmT \big] \big\} \Big]^\rmT.
\end{align}

\subsubsection{\textcolor{black}{First Stage-Direct Downlink Communication Channel Estimation}}
\textcolor{black}{The direct communication channel, $\bfd_j$, is estimated at the downlink $D_j$ in $\rmS_1$.
The input-output pairs for the proposed ELM \textcolor{black}{approach} are designed in a similar way as that at the ISAC BS.}

\textcolor{black}{\textbf{First Type of Input:}} With the received signals in \eqref{eq:r_jc}, the first type of input for the {proposed} ELM \textcolor{black}{approach} at the downlink $D_j$, $\bdzeta_{D_j}^{\rmS_1\rmI_1}$, is constructed by
\begin{align}
& \bdzeta_{D_j}^{\rmS_1\rmI_1}=\Big[\Re \big\{\big[\bfr_{j,1}^{\rmS_1},\ldots,\bfr_{j,C^{\rmS_1}}^{\rmS_1}\big] \big\}, \Im \big\{\big[\bfr_{j,1}^{\rmS_1},\ldots,\bfr_{j,C^{\rmS_1}}^{\rmS_1}\big] \big\}\Big]^\rmT.
\end{align}

\textcolor{black}{\textbf{Second Type of Input:}} \textcolor{black}{Based on the model in \eqref{eq:r_jc}}, the LS estimation of the BS-$D_j$ channel, $\bar\bfd_j$, is derived at the downlink $D_j$ by
\begin{align}
\bar\bfd_j = \mathbb E\big\{(\bfr_{j,c}^{\rmS_1} (\bfX^{\rmS_1})^\dag)^\rmH \big\} = \bfd_j + \mathbb E\big\{ (\tilde\bfw_{j,c}^{\rmS_1})^\rmH\big\}, \quad j\in\calN_1^J,
\label{eq:d_jbar}
\end{align}
where $\tilde{\bfw}_{j,c}^{\rmS_1}={\bfw}_{j,c}(\bfz_k^{\rmS_1})^\dag$.
Then, the second type \textcolor{black}{for the} input {of} the \textcolor{black}{proposed} ELM \textcolor{black}{approach} at the downlink $D_j$, $\bdzeta_{D_j}^{\rmS_1\rmI_2}$, is designed as
\begin{align}
\bdzeta_{D_j}^{\rmS_1\rmI_2}=\Big[\Re\{ \bar\bfd_j^\rmT \}, \Im\{ \bar\bfd_j^\rmT \} \Big]^\rmT.
\end{align}

\textcolor{black}{\textbf{Output:}} The output of the {proposed} ELM \textcolor{black}{approach} at the downlink $D_j$, $\bdgamma_{D_j}^{\rmS_1}$, is generated by the actual value of $\bfd_j$ as
\begin{align}
\bdgamma_{D_j}^{\rmS_1}=\Big[\Re\{ \bfd_j^\rmT \}, \Im\{ \bfd_j^\rmT \} \Big]^\rmT.
\end{align}

\subsubsection{\textcolor{black}{Second Stage-Reflected Uplink Communication Channel Estimation}}
\textcolor{black}{In $\rmS_2$, to estimate the reflected uplink communication channel $\bfB_k$ at the ISAC BS, the input-output pairs designs for the proposed ELM \textcolor{black}{approach} are presented as follows.}

\textcolor{black}{\textbf{First Type of Input:}} \textcolor{black}{By} employing the estimated direct SAC channels (i.e., $\hat\bfA$ and $\hat\bfb_k$) from $\rmS_1$ and received signals $\bfY_c^{\rmS_2}$ in \eqref{eq:Y_c_S2}, the first type of input for the \textcolor{black}{proposed} ELM \textcolor{black}{approach} at the ISAC BS, $\bdzeta_\rmBS^{\rmS_2\rmI_1}$, is designed as
\begin{align}
\bdzeta_\rmBS^{\rmS_2\rmI_1}
& = \Big[\Re\big\{\big[ {\rm{vec}}[\bfY_{C^{\rmS_1}+1}^{\rmS_2}, \bfY_{C^{\rmS_1}+2}^{\rmS_2}, \ldots, \bfY_{C^{\rmS_2}}^{\rmS_2}], {\rm{vec}}[\hat\bfA], \hat\bfb_1^\rmT,\hat\bfb_2^\rmT,\ldots,\hat\bfb_K^\rmT \big]\big\}, \notag\\
& \qquad \Im\big\{\big[ {\rm{vec}}[\bfY_{C^{\rmS_1}+1}^{\rmS_2}, \bfY_{C^{\rmS_1}+2}^{\rmS_2}, \ldots, \bfY_{C^{\rmS_2}}^{\rmS_2}], {\rm{vec}}[\hat\bfA], \hat\bfb_1^\rmT,\hat\bfb_2^\rmT,\ldots,\hat\bfb_K^\rmT \big]\big\}\Big]^{\rmT}.
\label{eq:zeta_S2I1_BS}
\end{align}

\textcolor{black}{\textbf{Second Type of Input:}} The second type {for the} input \textcolor{black}{of} the {proposed} ELM \textcolor{black}{approach} at the ISAC BS, $\bdzeta_\rmBS^{\rmS_2\rmI_2}$, relies on the LS estimates of the reflected $U_k$-IRS-BS channel that denoted by $\bar\bfB_k$.
To acquire $\bar\bfB_k$, the uplink reflected communication signal is firstly estimated as 
\begin{align}
\tilde\bfY_c^{\rmS_2}
& = \bfY_c^{\rmS_2} - \hat\bfA^\rmH\bfX^{\rmS_2} - \sum_{k\in\calN_1^K}\hat\bfb_k\bfz_k^{\rmS_2} \notag \\
& = \sum_{k\in\calN_1^K}\bfB_k\bfv_c^{\rmS_2}\bfz_k^{\rmS_2} + \tilde\bfN_c^{\rmS_2}, \quad c\in\calN_{C^{\rmS_1}+1}^{C^{\rmS_2}},
\label{eq:Y_ctilde_S2}
\end{align}
where $\tilde\bfN_c^{\rmS_2}= (\bfA^\rmH-\hat\bfA^\rmH)\bfX^{\rmS_2} + \sum_{k\in\calN_1^K}(\bfb_k - \hat\bfb_k)\bfz_k^{\rmS_2} + \bfN_c^{\rmS_2}$.
Then, by multiplying $(\bfz_k^{\rmS_2})^\dag$ with $\tilde\bfY_c^{\rmS_2}$ in \eqref{eq:Y_ctilde_S2}, the reflected signal from the uplink $U_k$ can be separated at the ISAC BS as $\bar\bfy_{k,c}^{\rmS_2}\in\bbC^{M\times 1}$, i.e.,
\begin{align}
\bar\bfy_{k,c}^{\rmS_2}
& = \tilde\bfY_c^{\rmS_2} (\bfz_k^{\rmS_2})^\dag  \notag\\
& = \bfB_k\bfv_c^{\rmS_2} + \bar\bfn_{k,c}^{\rmS_2}, \quad k\in\calN_1^K, \quad c\in\calN_{C^{\rmS_1}+1}^{C^{\rmS_2}}, 
\label{eq:y_bar_S2}
\end{align}
where $\bar\bfn_{k,c}^{\rmS_2} = \tilde\bfN_c^{\rmS_2}(\bfz_k^{\rmS_2})^\dag$.
Within the $(C^{\rmS_1}+1)$-th to the $C^{\rmS_2}$-th sub-frames, the matrix form of \eqref{eq:y_bar_S2} is
\begin{align}
\bar\bfY_k^{\rmS_2} = \bfB_k\bfV^{\rmS_2} + \bar\bfN_k^{\rmS_2}, \quad k\in\calN_1^K,
\end{align}
where $\bar\bfY_k^{\rmS_2}=[\bar\bfy_{k,C^{\rmS_1}+1}^{\rmS_2}, \bar\bfy_{k,C^{\rmS_1}+2}^{\rmS_2},\ldots, \bar\bfy_{k,C^{\rmS_2}}^{\rmS_2}]$ and $\bar\bfN_k^{\rmS_2} = [\bar\bfn_{k,C^{\rmS_1}+1}^{\rmS_2}, \bar\bfn_{k,C^{\rmS_1}+2}^{\rmS_2}, \ldots, $ $ \bar\bfn_{k,C^{\rmS_2}}^{\rmS_2}]$. 
Then, the LS estimate of the $U_k$-IRS-BS channel is given by 
\begin{align}
\bar\bfB_k = \bar\bfY_k^{\rmS_2} \big(\bfV^{\rmS_2}\big)^\dag, \quad k\in\calN_1^K.
\label{eq:B_k_bar}
\end{align}
With $\bar\bfB_k$ in \eqref{eq:B_k_bar}, $\bdzeta_\rmBS^{\rmS_2\rmI_2}$ is constructed by
\begin{align}
& \bdzeta_\rmBS^{\rmS_2\rmI_2}=\Big[\Re \big\{ {\rm{vec}}\big[\bar\bfB_1, \bar\bfB_2,\ldots, \bar\bfB_K\big] \big\}, \Im \big\{ {\rm{vec}}\big[\bar\bfB_1, \bar\bfB_2,\ldots, \bar\bfB_K\big] \big\}\Big]^\rmT.
\label{eq:zeta_S2I2_BS}
\end{align}

\textcolor{black}{\textbf{Output:}} For both inputs $\bdzeta_\rmBS^{\rmS_2\rmI_1}$ and $\bdzeta_\rmBS^{\rmS_2\rmI_2}$, the corresponding output of the {proposed} ELM \textcolor{black}{approach} at the ISAC BS, $\bdgamma_\rmBS^{\rmS_2}$, is formed by the actual values of the uplink reflected communication channels $\bfB_k$ as
\begin{align}
& \bdgamma_\rmBS^{\rmS_2}=\Big[\Re \big\{ {\rm{vec}}\big[\bfB_1, \bfB_2,\ldots, \bfB_K\big] \big\},  \Im \big\{ {\rm{vec}}\big[\bfB_1, \bfB_2,\ldots, \bfB_K\big] \big\}\Big]^\rmT.
\end{align}

\subsubsection{\textcolor{black}{Second Stage-Reflected Downlink Communication Channel Estimation}}
\textcolor{black}{The reflected downlink communication channel, $\bfD_j$, is estimated at the downlink $D_j$.
The corresponding input-output pairs designs for the proposed ELM \textcolor{black}{approach} are similar to that at the ISAC BS.}

\textcolor{black}{\textbf{First Type of Input:}} \textcolor{black}{The} first type \textcolor{black}{for the} input {of} the {proposed} ELM \textcolor{black}{approach}, $\bdzeta_{D_j}^{\rmS_2\rmI_1}$, is constructed by the estimated direct communication channel $\hat\bfd_j$ from $\rmS_1$ and received signals $\bfr_{j,c}^{\rmS_2}$ in \eqref{eq:r_jc_S2} as
\begin{align}
 \bdzeta_{D_j}^{\rmS_2\rmI_1}
& = \Big[\Re\big\{\big[\bfr_{j,C^{\rmS_1}+1}^{\rmS_2}, \bfr_{j,C^{\rmS_1}+2}^{\rmS_2}, \ldots, \bfr_{C^{\rmS_2}}^{\rmS_2}, \hat\bfd_j^\rmT \big]\big\}, \notag\\
& \qquad \Im\big\{\big[ \bfr_{j,C^{\rmS_1}+1}^{\rmS_2}, \bfr_{j,C^{\rmS_1}+2}^{\rmS_2}, \ldots, \bfr_{C^{\rmS_2}}^{\rmS_2}, \hat\bfd_j^\rmT \big]\big\}\Big]^{\rmT}.
\end{align}

\textcolor{black}{\textbf{Second Type of Input:}} The second type \textcolor{black}{for the} input {of} the {proposed} ELM \textcolor{black}{approach} at the downlink $D_j$, $\bdzeta_{D_j}^{\rmS_2\rmI_2}$, is built on the LS estimate of the reflected BS-IRS-$D_j$ channel, defined as $\bar\bfD_j$.
With $\hat\bfd_j$ from $\rmS_1$ and $\bfr_{j,c}^{\rmS_2}$ in \eqref{eq:r_jc_S2}, the rough estimation of the downlink reflected communication signal is firstly derived by
\begin{align}
\tilde\bfr_{j,c}^{\rmS_2}
& = \bfr_{j,c}^{\rmS_2} - \hat\bfd_j^\rmH\bfX^{\rmS_2}  \notag\\
& = (\bfv_c^{\rmS_2})^\rmH \bfD_j^\rmH \bfX^{\rmS_2} + \tilde\bfw_{j,c}^{\rmS_2}, \quad j\in\calN_1^J, c\in\calN_{C^{\rmS_1}+1}^{C^{\rmS_2}},
\label{eq:r_tilde_S2}
\end{align}
where $\tilde\bfw_{j,c}^{\rmS_2}=(\bfd_j - \hat\bfd_j)\bfX^{\rmS_2} + \bfw_{j,c}^{\rmS_2}$.
Then, by separating $\bfX^{\rmS_2}$ from $\tilde\bfr_{j,c}^{\rmS_2}$ in \eqref{eq:r_tilde_S2}, the resulted $\bar\bfr_{j,c}^{\rmS_2}\in\bbC^{1\times M}$ is expressed as
\begin{align}
\bar\bfr_{j,c}^{\rmS_2}
& = \tilde\bfr_{j,c}^{\rmS_2} (\bfX^{\rmS_2})^\dag \notag\\
& = (\bfv_c^{\rmS_2})^\rmH\bfD_j^\rmH + \bar\bfw_{j,c}^{\rmS_2}, \quad j\in\calN_1^J, \quad c\in\calN_{C^{\rmS_1}+1}^{C^{\rmS_2}},
\label{eq:r_bar_S2}
\end{align}
where $\bar\bfw_{j,c}^{\rmS_2} = \tilde\bfw_{j,c}^{\rmS_2}(\bfX^{\rmS_2})^\dag$.
The matrix form of \eqref{eq:r_bar_S2} is written as
\begin{align}
\bar\bfR_j^{\rmS_2} = \big(\bfV^{\rmS_2}\big)^\rmH \bfD_j^\rmH + \bar\bfW_j^{\rmS_2}, \quad j\in\calN_1^J,
\end{align}
where $\bar\bfR_j^{\rmS_2}=[(\bar\bfr_{j,C^{\rmS_1}+1}^{\rmS_2})^\rmT,(\bar\bfr_{j,C^{\rmS_1}+2}^{\rmS_2})^\rmT, \ldots,(\bar\bfr_{j,C^{\rmS_2}}^{\rmS_2})^\rmT]^\rmT$ and $\bar\bfW_j^{\rmS_2} = [(\bar\bfw_{j,C^{\rmS_1}+1}^{\rmS_2})^\rmT, (\bar\bfw_{j,C^{\rmS_1}+2}^{\rmS_2})^\rmT,\ldots,$ $(\bar\bfw_{j,C^{\rmS_2}}^{\rmS_2})^\rmT]^\rmT$.
Consequently, the LS estimate of the BS-IRS-$D_j$ channel is given by
\begin{align}
\bar\bfD_j = \big(\bar\bfR_j^{\rmS_2} \big)^\rmH \big(\bfV^{\rmS_2}\big)^\dag, \quad j\in\calN_1^J.
\label{eq:D_jbar}
\end{align}
By using $\bar\bfD_j$ in \eqref{eq:D_jbar}, $\bdzeta_{D_j}^{\rmS_2\rmI_2}$ is generated as
\begin{align}
\bdzeta_{D_j}^{\rmS_2\rmI_2}=\Big[\Re \big\{ {\rm{vec}}\big[\bar\bfD_j\big] \big\}, \Im \big\{ {\rm{vec}}\big[ \bar\bfD_j \big] \big\}\Big]^\rmT.
\end{align}

\textcolor{black}{\textbf{Output:}} Regarding the above two types of \textcolor{black}{the} inputs (i.e, $\bdzeta_{D_j}^{\rmS_2\rmI_1}$ and $\bdzeta_{D_j}^{\rmS_2\rmI_2}$), the output of the \textcolor{black}{proposed} ELM at the downlink $D_j$, $\bdgamma_{D_j}^{\rmS_2}$, is formed by the actual value of the reflected communication channel $\bfD_j$ as
\begin{align}
\bdgamma_{D_j}^{\rmS_2}=\Big[\Re \big\{ {\rm{vec}}\big[\bfD_j\big] \big\}, \Im \big\{ {\rm{vec}}\big[ \bfD_j \big] \big\}\Big]^\rmT.
\end{align}

\begin{figure}
\centering
\subfigure[ ]
{\begin{minipage}[b]{0.5\textwidth}
\includegraphics[width=1\textwidth]{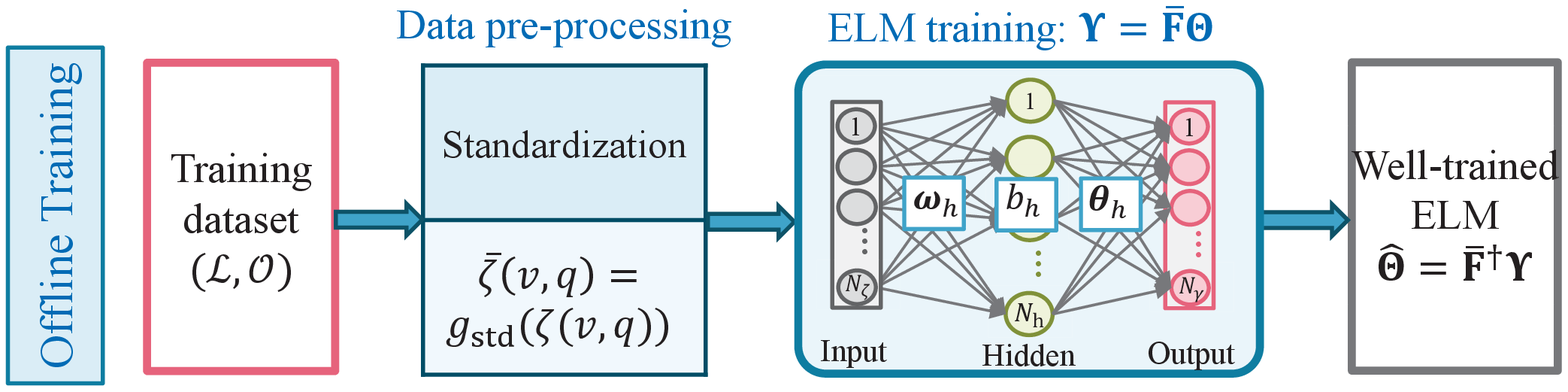}
\end{minipage}}

\subfigure[ ]
{\begin{minipage}[b]{0.6\textwidth}
\includegraphics[width=1\textwidth]{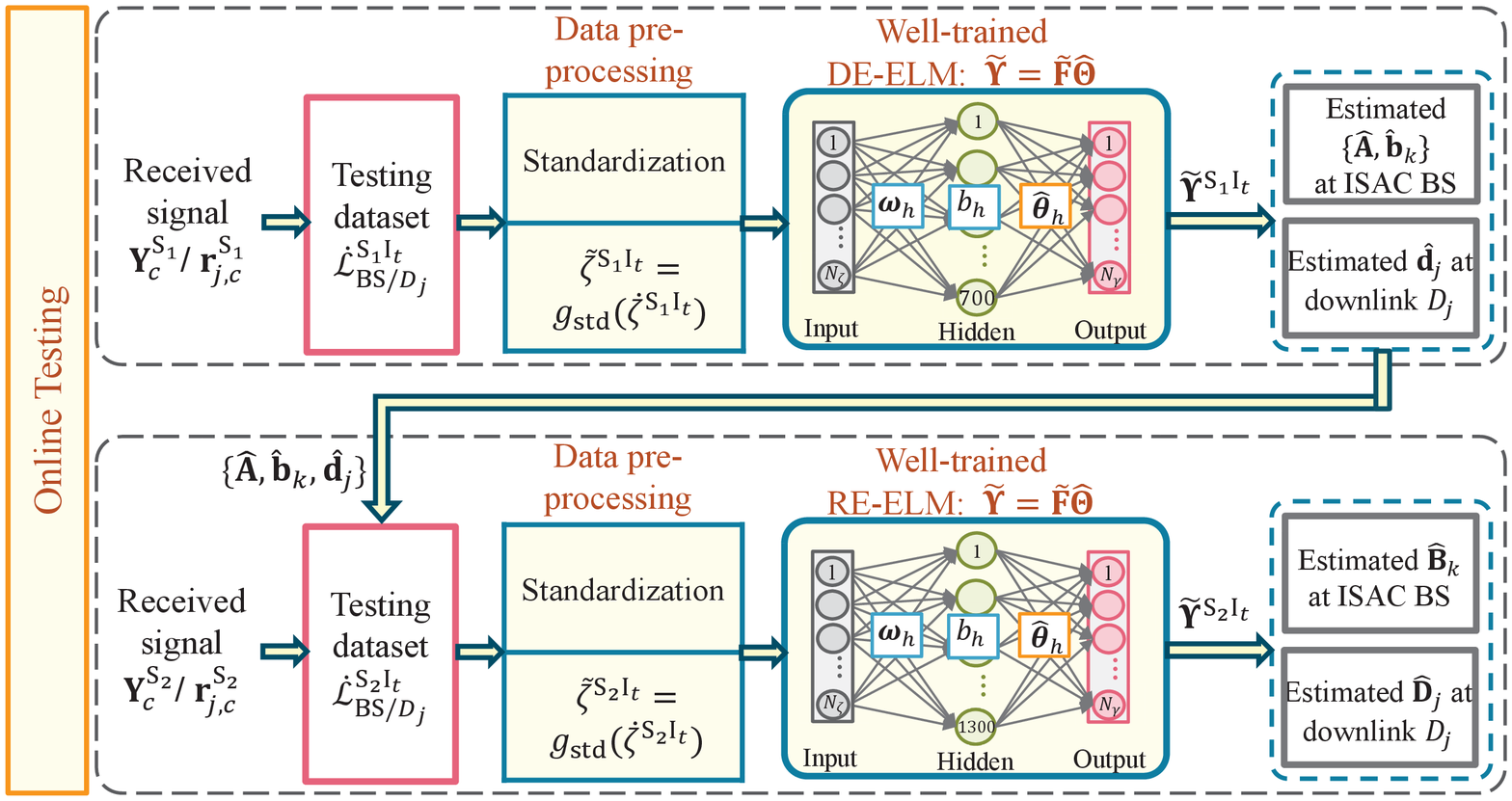}
\end{minipage}}
\caption{The proposed ELM-based \textcolor{black}{NN} framework: (a) Offline training, (b) Online testing.}
\label{fig:ELM_Diag}
\end{figure}

\subsection{Proposed ELM-based Estimation Framework}\label{sec:DL-CE-appraoch}
In light of the proposed two-stage approach, this section proposes an ELM-based \textcolor{black}{NN} framework to estimate the SAC channels.
\textcolor{black}{The training dataset is constructed by the designed input-output pairs to train the proposed ELM.
The overall proposed ELM framework is illustrated in Fig. \ref{fig:ELM_Diag}, involving the offline training and online testing phases.}

\subsubsection{Training Dataset Generation}\label{sec:dataset}
Based on the designed input-output pairs in the two estimation stages, the training datasets of the ELM are generated as follows:
At the ISAC BS, the training dataset that corresponds to the $t$-th type of input-output pair in the $\ell$-th stage is given by
\begin{align}
& \big(\calL_\rmBS^{\rmS_\ell\rmI_t},\calO_\rmBS^{\rmS_\ell}\big)
 = \Big\{ \Big(\bdzeta_{\rmBS}^{\rmS_\ell\rmI_t}(1,1),\bdgamma_{\rmBS}^{\rmS_\ell}(1)\Big),
\Big(\bdzeta_{\rmBS}^{\rmS_\ell\rmI_t}(1,2),\bdgamma_{\rmBS}^{\rmS_\ell}(1)\Big), \notag\\
& \qquad \ldots,\Big(\bdzeta_{\rmBS}^{\rmS_\ell\rmI_t}(1,Q), \bdgamma_{\rmBS}^{\rmS_\ell}(1)\Big),
\Big(\bdzeta_{\rmBS}^{\rmS_\ell\rmI_t}(2,1),\bdgamma_{\rmBS}^{\rmS_\ell}(2)\Big), \ldots, \notag \\
& \qquad \Big(\bdzeta_{\rmBS}^{\rmS_\ell\rmI_t}(V,Q),\bdgamma_{\rmBS}^{\rmS_\ell}(V)\Big)  \Big\}, \quad \ell\in\calN_1^2, \quad t\in\calN_1^2,
\label{eq:L_cal_BS}
\end{align}
where \textcolor{black}{$\big(\bdzeta_{\rmBS}^{\rmS_\ell\rmI_t}(v,q),\bdgamma_{\rmBS}^{\rmS_\ell}(v)\big)$} denotes the $(v,q)$-th, $v\in\calN_1^V$, $q\in\calN_1^Q$, training sample.
For the convenience of understanding, the detailed generation process of the training dataset $\big(\calL_{\rmBS}^{\rmS_1\rmI_1},\calO_{\rmBS}^{\rmS_1}\big)$ is introduced here.
In light of the data augmentation concept, the training samples can be enriched by creating synthetic samples from the existing ones \cite{ref:data-augment}.
Then, the inputs of the {proposed} ELM in {that} dataset are generated by using $V$ existing received signals at the ISAC BS (i.e., SAC signals in \eqref{eq:Y_c}), and $Q-1$ copies of the $v$-th are created by inducing synthetic noise to the existing SAC channels under $\text{SNR}_\text{ch} =\frac{{\cal P}_\text{ch} }{\sigma_\text{ch}^2}$.
\textcolor{black}{Here, ${\cal P}_\text{ch}$ represents the power of the existing channel, and the synthetic noise follows ${\cal CN}(0,\sigma_\text{ch}^2)$ with zero-mean and variance of $\sigma_\text{ch}^2$.}
\textcolor{black}{Therefore, {the} synthetic channels can be used to represent the SAC signals, and then, create the synthetic samples (i.e., $\big(\bdzeta_\rmBS^{\rmS_1\rmI_1}(v,q),\bdgamma_\rmBS^{\rmS_1}(v)\big)$, $v\in\calN_{1}^V$, $q\in\calN_{2}^Q$).}
Combined with the existing samples (i.e., $\big(\bdzeta_\rmBS^{\rmS_1\rmI_1}(v,1),\bdgamma^{\rmS_1}(v)\big)$, $v\in\calN_{1}^V$), the dataset $\big(\calL_\rmBS^{\rmS_1\rmI_1},\calO_\rmBS^{\rmS_1}\big)$ is \textcolor{black}{finally} generated.
It is worth mentioning that the enrichment of the training samples aims to improve the channel estimation performance of the ELM.
In addition, the training datasets for the \textcolor{black}{proposed} ELM at the downlink UEs (i.e., $\big(\calL_{D_j}^{\rmS_\ell\rmI_t},\calO_{D_j}^{\rmS_\ell}\big)$, $j\in\calN_1^J$) are generated similarly as in \eqref{eq:L_cal_BS}.

\textcolor{black}{\subsubsection{Brief Introduction of ELM}\label{sec:ELM-introduction}
In essence, ELM is structured as an FNN that consists of the input, single hidden, and output layers.
However, the learning of the ELM parameters is different. To illustrate this, let $N_\zeta$, $N_\rmh$, and $N_\gamma$ denote the number of neurons adopted in the above three layers, respectively.
To facilitate the understanding of the ELM training procedure, the superscript (i.e., $\rmS_\ell\rmI_t$) and subscript (i.e., $\rmBS$ and $D_j$) of the variables are omitted.
Given the training dataset $\big(\calL,\calO\big)$ in \eqref{eq:L_cal_BS}, the data pre-processing on this dataset is firstly required to enhance the learning capacity of the ELM. 
Given the $(v,q)$-th pre-processed input-output pair as $(\bar\bdzeta(v,q),\bdgamma(v))$, the ELM is mathematically modeled as} 
\begin{align}
\bdgamma(v)=\sum_{h\in\calN_1^{N_\rmh}} \bdtheta_h f\big(\bdomega_h\cdot\bar\bdzeta(v,q)+b_h\big), v\in\calN_1^V, q\in\calN_1^Q,
\label{eq:gama_v}
\end{align}
where $\bdomega_h=[\omega_{h,1},\omega_{h,2},\ldots,\omega_{h,N_\zeta}]^\rmT\in\bbC^{N_\zeta\times 1}$, $b_h$, and $f(\cdot)$  represent the input weight vector that connects the $h$-th hidden neuron and $N_\zeta$ input neurons, bias of the $h$-th hidden neuron, and activation function, respectively.
Define $\bdtheta_h=[\theta_{h,1},\theta_{h,2},\ldots,\theta_{h,N_\gamma}]^\rmT\in\bbC^{N_\gamma\times 1}$ as the output weight vector between the $h$-th hidden neuron and $N_\gamma$ output neurons.
Accordingly, for $N_\rmtr=VQ$ input-output pairs, the matrix form of \eqref{eq:gama_v} can be compactly formulated as
\begin{align}
\bold\Upsilon = \bar\bfF\bold\Theta,
\label{eq:Gamma}
\end{align}
where
\begin{align}
&\bar\bfF =  \left[
\begin{array}{cccc}
f\big(\bdomega_1\bar\bdzeta(1,1)+b_1\big) & \cdots & f\big(\bdomega_{N_\rmh}\bar\bdzeta(1,1)+b_{N_\rmh}\big) \\
\vdots & \ddots & \vdots \\
f\big(\bdomega_1\bar\bdzeta(V,Q)+b_1\big) & \cdots & f\big(\bdomega_{N_\rmh}\bar\bdzeta(V,Q)+b_{N_\rmh}\big)
\end{array}
\right] 
\label{eq:F_output}
\end{align}
\begin{align}
\bold\Theta =[\bdtheta_1,\bdtheta_2,\ldots,\bdtheta_{N_\rmh}]^\rmT\in\bbR^{N_\rmh\times N_\gamma}, \notag
\end{align}
\begin{align}
\bold\Upsilon =[\bdgamma(1),\bdgamma(2),\ldots,\bdgamma({N_\rmtr})]^\rmT\in\bbR^{N_\rmtr \times N_\gamma}.
\end{align}
Here, $\bar\bfF$ denotes the output matrix of the hidden layer \cite{ref:ELM1}.
Different from the traditional gradient-based learning algorithms {used in traditional FNN}, the ELM provides a more efficient solution to train the NN parameters in \eqref{eq:Gamma} (i.e., $\bdomega_h$ and $b_h$ in $\bar\bfF$, and $\bold\Theta$).
\textcolor{black}{It has been proved in \cite{ref:ELM1} that if the activation function $f(\cdot)$ is infinitely differentiable, the parameters $\bdomega_h$ and $b_h$ in $\bar\bfF$ can be initialized randomly and do not need to be tuned during the training process.}
\textcolor{black}{This proof enables \eqref{eq:Gamma} to become a linear system when $\bar\bfF$ remains constant with the unchanged $\bdomega_h$ and $b_h$.
In such a case, the designed ELM is only required to update the parameter $\bold\Theta$ by minimizing the loss function as}
\begin{align}
\hat{\bold\Theta} = \min_\bold\Theta|| \bar\bfF\bold\Theta - \bold\Upsilon||_{2}.
\label{eq:Theta_hat_min}
\end{align}
By finding the LS solution of \eqref{eq:Theta_hat_min}, $\hat{\bold\Theta}$ is determined by
\begin{align}
\hat{\bold\Theta}=\bar\bfF^\dag\boldsymbol\Upsilon.
\label{eq:Theta_hat}
\end{align}
\textcolor{black}{According to the above ELM network training procedure,} the advantages of the proposed ELM-based \textcolor{black}{NN} estimation framework are attributed to three aspects as follows:
First, the ELM significantly \textcolor{black}{promotes} the learning speed and reduces the NN parameters compared to the gradient-based learning algorithms.
\textcolor{black}{It can be noted that when $\bar\bfF$ is unchanged during the training process, the linear relationship in \eqref{eq:Gamma} always holds.
Therefore, unlike the common understanding that most of the NN parameters need to be adjusted through backpropagation algorithms, the ELM only needs to determine the NN weights $\bold\Theta$ by performing the matrix-inversion.}
Second, the ELM can analytically reach the smallest training error through the derived solution in \eqref{eq:Theta_hat}.
Note that most of the learning algorithms cannot achieve this smallest training error since they may get stuck in {a} local minimum, suffer from overfitting, or choose an improper learning rate \cite{ref:ELM1}.
Third, the generalization ability of {the} ELM is considerable.
This makes the ELM applicable for the practical channel estimation issue under various SNR conditions.

\begin{table}[t] 
\newcommand{\tabincell}[2]{\begin{tabular}{@{}#1@{}}#2\end{tabular}}
\centering
\caption{Hyperparameters of DE-ELM and RE-ELM.} \label{table:ELM}
\begin{tabular}{p{1.2cm}<{\centering}| p{2cm}<{\centering}| p{5cm}<{\centering}| p{2.8cm}<{\centering}}
\hline
                            & {\bf Layer type} & {\bf Tensor size} & {\bf Activation function} \\
\hline
\hline
\multirow{7}{*}{\bf{DE-ELM}} & \multirow{2}{*}{Input (BS)} & ${\rmS_1\rmI_1}: 2MP^{\rmS_1}C^{\rmS_1}\,\,\,$ &
                                               \multirow{4}{*}{-} \\
                            &                                       & ${\rmS_1\rmI_2}: 2M(M+K)$  & \\ \cline{2-3}
                            & \multirow{2}{*}{Input ($D_j$)}        & ${\rmS_1\rmI_1}: 2P^{\rmS_1}C^{\rmS_1}\quad\,\,\,$ &  \\
                            &                         & ${\rmS_1\rmI_2}: 2M\qquad\quad\,\,\,\,\,$  &        \\ \cline{2-4}
                            & Hidden                  & $700$                               & {\textit{linear}} \\ \cline{2-4}
                            & Output (BS)             & ${\rmS_1}: 2M(M+K)$  & \multirow{2}{*}{-}       \\ \cline{2-3}
                            & Output ($D_j$)          & ${\rmS_1}: 2M\qquad\qquad\,$        &    \\ \cline{1-4}
\multirow{7}{*}{\bf {RE-ELM}} & \multirow{2}{*}{Input (BS)}   & ${\rmS_2\rmI_1}: 2M(P^{\rmS_2}(C^{\rmS_2}-C^{\rmS_1})+M+K)$  & \multirow{6}{*}{-}\\
                            &                & ${\rmS_2\rmI_2}: 2MLK\qquad\qquad\qquad\qquad\quad\,\,\,\,$    & \\ \cline{2-3}
                            & \multirow{2}{*}{Input ($D_j$)}      & ${\rmS_2\rmI_1}: 2(P^{\rmS_2}(C^{\rmS_2}-C^{\rmS_1})+M)$    & \\
                            &                & ${\rmS_2\rmI_2}: 2ML\qquad\qquad\qquad\qquad\,\,\,$    &  \\ \cline{2-4}
                            & Hidden                  & $1300$                &{{\textit{sigmoid}}}  \\ \cline{2-4}
                            & Output (BS)             & ${\rmS_2}: 2MLK$     & \multirow{2}{*}{-}          \\ \cline{2-3}
                            & Output ($D_j$)          & ${\rmS_2}: 2ML\quad$ &                             \\ \cline{1-4}
\hline
\end{tabular}
\end{table} 

\subsubsection{Offline Training}

For the SAC channels estimation in both $\rmS_1$ and $\rmS_2$, the offline training phase consists of the input standardization and ELM training procedures, as shown in Fig. \ref{fig:ELM_Diag}(a).
The ELM networks in the developed \textcolor{black}{NN} framework are trained by the standardized samples to attain the well-trained ELMs.
Two different ELM structures are devised to build {the proposed} \textcolor{black}{NN} framework.
The first ELM structure is employed to estimate the direct SAC channels in $\rmS_1$, namely direct estimation ELM (DE-ELM), whereas the other one is designed for the reflected communication channels estimation in $\rmS_2$, referred to as reflected estimation ELM (RE-ELM).

\textcolor{black}{The offline training at the ISAC BS and downlink UEs performs similarly in the two estimation stages, as follows.}
Given the training dataset $\big(\calL,\calO\big)$ in \eqref{eq:L_cal_BS}, the $(v,q)$-th input $\bdzeta(v,q)$ in {the} dataset is firstly standardized as $\bar\bdzeta(v,q)$, denoted by
\begin{align}
\bar\bdzeta(v,q)
& = g_{\rm std}\big(\bdzeta(v,q)\big) \notag\\
& =\frac{\bdzeta(v,q)-{\mathbb E}\{\bdzeta(v,q)\} }{ {\mathbb D}\{\bdzeta(v,q)\} }, \quad v\in\calN_1^V,  q\in\calN_1^Q,
\label{eq:f_std}
\end{align}
where $g_{\rm std}(\cdot)$ represents the standardization operator.
Then, the \textcolor{black}{proposed} ELM network is adopted to learn the mapping function of the standardized input-output pair $\big(\bar\bdzeta(v,q),\bdgamma(v)\big)$.
\textcolor{black}{The ELM training procedure follows the introduction in Section \ref{sec:ELM-introduction}.}

\newlength\myindent 
\setlength\myindent{2em}
\newcommand\bindent{
    \begingroup
    \setlength{\itemindent}{\myindent}
    \addtolength{\algorithmicindent}{\myindent} }
\newcommand\eindent{\endgroup}

\begin{algorithm}[!t]
\caption{ELM-based Channel Estimation Algorithm.}
{\bf Offline training:}

\begin{algorithmic}[1]  
  \FOR {$\rmS_\ell$, $\ell\in\calN_1^2$,}

  \FOR {$\rmI_t$, $t\in\calN_1^2$,}
    \STATE {\bf Generate} $(\calL_\rmBS^{\rmS_\ell\rmI_k},\calO_\rmBS^{\rmS_\ell})$ at the ISAC BS and $(\calL_{D_j}^{\rmS_\ell\rmI_k},\calO_{D_j}^{\rmS_\ell})$ at the downlink $D_j$ according to Sections \ref{sec:IO-design} and \ref{sec:dataset};
    \STATE {\bf Pre-process} $(\calL_\rmBS^{\rmS_\ell\rmI_k},\calO_\rmBS^{\rmS_\ell})$ and $(\calL_{D_j}^{\rmS_\ell\rmI_k},\calO_{D_j}^{\rmS_\ell})$;
    \STATE {\bf Input} the standardized samples $(\bar\bdzeta_\rmBS^{\rmS_\ell\rmI_k}(v,q), $ $\bdgamma_\rmBS^{\rmS_\ell}(q))$ and $(\bar\bdzeta_{D_j}^{\rmS_\ell\rmI_k}(v,q), \bdgamma_{D_j}^{\rmS_\ell}(q))$, respectively, to the designed ELM in \eqref{eq:Gamma};
    \STATE {\bf Initialize} $\{\bdomega_{h},b_{h}\}$, $h\in\calN_1^{N_\rmh}$ randomly;
    \STATE {\bf Compute} $\bar\bfF_{\rmBS/D_j}$ according to \eqref{eq:F_output};
    \STATE {\bf Compute} $\hat{\bold\Theta}_{\rmBS/D_j}$ according to \eqref{eq:Theta_hat};
    \STATE {\bf Output} the trained ELM with $\hat{\bold\Theta}_\rmBS$ at the ISAC BS and trained ELM with $\hat{\bold\Theta}_{D_j}$ at the downlink $D_j$.

  \ENDFOR

  \ENDFOR
\end{algorithmic}

{\bf Online testing:}
\begin{algorithmic}[1]  

  \STATE {\bf Input:} Testing dataset $\dot\calL_\rmBS^{\rmS_\ell\rmI_t}$ and $\dot\calL_{D_j}^{\rmS_\ell\rmI_t}$, $\ell\in\calN_1^2$, $t\in\calN_1^2$;
  \begin{ALC@g}
    \STATE {\bf Estimate} $\{\bfA,\bfb_k\}$, $k\in\calN_1^K$ at the ISAC BS and $\bfd_j$ at the downlink $D_j$ by using the trained DE-ELM according to \eqref{eq:Gamma_tilde}; 
    \STATE {\bf Estimate} $\bfB_k$, $k\in\calN_1^K$ at the ISAC BS and $\bfD_j$ at the downlink $D_j$ by using the trained RE-ELM according to \eqref{eq:Gamma_tilde}; 
  \end{ALC@g}
  \STATE {\bf Output:} Estimated channels $\{\hat\bfA, \hat\bfb_k, \hat\bfd_j, \hat\bfB_k, \hat\bfD_j\}$, $k\in\calN_1^K$, $j\in\calN_1^J$.

\end{algorithmic}
\label{algo:DL-CE}
\end{algorithm}

\textcolor{black}{In the developed \textcolor{black}{NN} framework, the hyperparameters of the DE-ELM and RE-ELM are carefully designed regarding the different estimation tasks.
For the direct channel estimation,} the hidden layer of the DE-ELM is composed of $N_\rmh=700$ hidden neurons with \textit{linear} activation functions.
Since the reflected channel estimation is more complicated than the direct one, the hidden layer of the RE-ELM increases to $N_\rmh=1300$ hidden neurons with \textit{sigmoid} activation functions, enhancing its feature extraction ability.
\textcolor{black}{To further reduce the correlation of the training samples and improve the learning ability of the RE-ELM, a standardization operator is added before the activation function of the hidden layer in the RE-ELM \cite{ref:ELM-OFDM-CE}.
Since the {\textit{sigmoid}} activation function returns an output value in the range of $(0,1)$ and prevents the activation value from vanishing, it is an excellent choice for the RE-ELM hidden layer.}
The detailed hyperparameters of the DE-ELM and RE-ELM are summarized in Table \ref{table:ELM}.

\subsubsection{Online Testing}
The online testing phase for the two estimation stages is depicted in Fig. \ref{fig:ELM_Diag}(b).
The input $\dot{\bdzeta}$ in the testing dataset $\dot\calL$ is firstly standardized as $\tilde\bdzeta = g_{\rm std}\big(\dot{\bdzeta}\big)$. 
By substituting $\tilde\bdzeta$ and the initial values of $\{\bdomega_h,b_h\}$ into \eqref{eq:F_output}, the output matrix of the hidden layer is obtained as $\tilde\bfF$.
Then, with the determined hyperparameter $\hat{\bold\Theta}$ in \eqref{eq:Theta_hat}, the output of the trained ELM, $\tilde{\boldsymbol\Upsilon}$, is denoted by
\begin{align}
\tilde{\boldsymbol\Upsilon}=\tilde\bfF\hat{\bold\Theta}.
\label{eq:Gamma_tilde}
\end{align}
After that, the direct SAC and reflected communication channels are successively estimated as $\{\hat\bfA,\hat\bfb_k,\hat\bfd_j\}$ and $\{\hat\bfB_k,\hat\bfD_j\}$ by performing mathematical operation on $\tilde{\boldsymbol\Upsilon}$ (i.e., add up the real and imaginary parts in $\tilde{\boldsymbol\Upsilon}$ to attain the complex numbers).
Algorithm \ref{algo:DL-CE} summarizes the procedure of the proposed ELM-based channel estimation algorithm.

\section{Complexity Analysis}\label{sec:complexity}
In this section, the computational complexity of the inputs generation and ELM-based online testing for the proposed \textcolor{black}{NN} estimation approach is discussed.
The complexity is measured by the required number of real additions and multiplications.

\subsection{Complexity of Inputs Generation}
The first type of inputs (i.e., $\bdzeta_\rmBS^{\rmS_1\rmI_1}$, $\bdzeta_{D_j}^{\rmS_1\rmI_1}$, $\bdzeta_\rmBS^{\rmS_2\rmI_1}$, and $\bdzeta_{D_j}^{\rmS_2\rmI_1}$) are directly constructed by the received signals without a need of signal pre-processing.
Thus, there is no additional complexity for their inputs generation.

In $\rmS_1$, for generating the second type of input at the ISAC BS (i.e., $\bdzeta_\rmBS^{\rmS_1\rmI_2}$), the computational complexity \textcolor{black}{can be calculated for} \eqref{eq:A_bar} and \eqref{eq:b_bar}.
In \eqref{eq:A_bar}, the required number of real additions and multiplications are $\calA^{\bar{\rm A}}=\frac 2 3 C^{\rmS_1}M(18MP^{\rmS_1}+3M^2-3P^{\rmS_1}+3M-1)-2M^2$ and $\calM^{\bar{\rm A}}=\frac 1 3 C^{\rmS_1}M(36MP^{\rmS_1}+4M^2+15M-1)+2M^2+1$, respectively.
For estimating the reflected channel of each uplink UE, the cost of \eqref{eq:b_bar} is $\calA^{\bar{\rm b}_k}=\frac 2 3 C^{\rmS_1}(6MP^{\rmS_1}+9P^{\rmS_1}-3M+2)-2M$ real additions and $\calM^{\bar{\rm b}_k}=2C^{\rmS_1}(2MP^{\rmS_1}+4P^{\rmS_1}+3)+2(M+1)$ real multiplications.
Thus, the number of real additions and multiplications, $\calA_\rmBS^{\rmS_1\rmI_2}$ and $\calM_\rmBS^{\rmS_1\rmI_2}$, required to construct the input $\bdzeta_\rmBS^{\rmS_1\rmI_2}$ are \textcolor{black}{respectively} given by
\begin{align}
\calA_\rmBS^{\rmS_1\rmI_2}
& = \calA^{\bar{\rm A}} + K\calA^{\bar{\rm b}_k} \notag \\
& = \frac 2 3 MC^{\rmS_1}(18MP^{\rmS_1}+3M^2-3P^{\rmS_1}+3M-1) \notag\\
& + \frac 2 3 KC^{\rmS_1}(6MP^{\rmS_1}+9P^{\rmS_1}-3M+2) -2M(K+M),
\end{align}
and
\begin{align}
\calM_\rmBS^{\rmS_1\rmI_2}
& = \calM^{\bar{\rm A}} + K\calM^{\bar{\rm b}_k} \notag \\
& = \frac 1 3 MC^{\rmS_1}(36MP^{\rmS_1}+4M^2+15M-1)  + 2KC^{\rmS_1}(2MP^{\rmS_1} +4P^{\rmS_1}+3) \notag \\
& \qquad + 2M(K+M)+K+1.
\end{align}

The second type of input at the downlink $D_j$ (i.e., $\bdzeta_{D_j}^{\rmS_1\rmI_2}$) is constructed by the LS estimation results in \eqref{eq:d_jbar}.
Hence, the number of real additions and multiplications, $\calA_{D_j}^{\rmS_1\rmI_2}$ and $\calM_{D_j}^{\rmS_1\rmI_2}$, required to generate the input $\bdzeta_{D_j}^{\rmS_1\rmI_2}$ are respectively written as
\begin{align}
\calA_{D_j}^{\rmS_1\rmI_2}
& = \frac 2 3 MC^{\rmS_1}(12MP^{\rmS_1}+3M^2 +3P^{\rmS_1}-1)-2M,
\end{align}
and
\begin{align}
\calM_{D_j}^{\rmS_1\rmI_2}
& = \frac 1 3 MC^{\rmS_1}(24MP^{\rmS_1}+4M^2+12P^{\rmS_1}  +15M-1)+2M+1.
\end{align}

In $\rmS_2$, for generating the second type of input at the ISAC BS (i.e., $\bdzeta_\rmBS^{\rmS_2\rmI_2}$), the complexity comes from \eqref{eq:Y_ctilde_S2}, \eqref{eq:y_bar_S2}, and \eqref{eq:B_k_bar}.
The cost of \eqref{eq:Y_ctilde_S2} in each sub-frame is $\tilde\calA_\rmBS^{\rmS_2\rmI_2}=2MP^{\rmS_2}(K+M)$ real additions and $\tilde\calM_\rmBS^{\rmS_2\rmI_2}=4MP^{\rmS_2}(K+M)$ real multiplications.
\textcolor{black}{According} to \cite{ref:matrix-inverse}, the inverse calculation of an $m\times m$ complex matrix costs $\calA_{\rm inv}^m = \frac 2 3m(3m^2+3m-1)$ real additions and $\calM_{\rm inv}^m = \frac 1 3 m(4m^2 + 15m - 1)$ real multiplications.
On this basis, for the pseudoinverse calculation of an $a\times b$ complex matrix, the required number of real additions and multiplications are $\calA_{\rm pinv}^{(a,b)} = 8a^2b-2a(a+b)+\calA_{\rm inv}^a$ and $\calM_{\rm pinv}^{(a,b)} = 8a^2b+\calM_{\rm inv}^a$, respectively.
Then, for estimating the reflected channel of each uplink UE, the cost of \eqref{eq:y_bar_S2} in each sub-frame is $\bar\calA_\rmBS^{\rmS_2\rmI_2}$ $=2M(2P^{\rmS_2}-1)+\calA_{\rm pinv}^{(1,P^{\rmS_2})}$ and $\bar\calM_\rmBS^{\rmS_2\rmI_2}=4MP^{\rmS_2} $ $ +\calM_{\rm pinv}^{(1,P^{\rmS_2})}$ real additions and multiplications, respectively.
The LS estimator in \eqref{eq:B_k_bar} requires $\calA^{\bar\rmB_k}=2ML(2C^{\rmS_2}-2C^{\rmS_1}-1)+\calA_{\rm pinv}^{(L,(C^{\rmS_2}-C^{\rmS_1}))}$ real additions and $\calM^{\bar\rmB_k}=4ML(C^{\rmS_2}-C^{\rmS_1})+\calM_{\rm pinv}^{(L,(C^{\rmS_2}-C^{\rmS_1}))}$ real multiplications.
Therefore, the number of real additions and multiplications, $\calA_\rmBS^{\rmS_2\rmI_2}$ and $\calM_\rmBS^{\rmS_2\rmI_2}$, required to construct the input $\bdzeta_\rmBS^{\rmS_2\rmI_2}$ are \textcolor{black}{respectively} given by
\begin{align}
\calA_\rmBS^{\rmS_2\rmI_2}
& = (C^{\rmS_2}-C^{\rmS_1})(\tilde\calA_\rmBS^{\rmS_2\rmI_1} + K\bar\calA_\rmBS^{\rmS_2\rmI_1}) + K\calA^{\bar\rmB_k} \notag \\
& = \frac 2 3K(C^{\rmS_2}-C^{\rmS_1})(6ML+12L^2-3L-3M+2) \notag \\
& \qquad + 2P^{\rmS_2}(C^{\rmS_2}-C^{\rmS_1})(3MK+M+3K)  + \frac 2 3KL(3L^2-3M-1),
\end{align}
and
\begin{align}
\calM_\rmBS^{\rmS_2\rmI_2}
& = (C^{\rmS_2}-C^{\rmS_1})(\tilde\calM_\rmBS^{\rmS_2\rmI_1} + K\bar\calM_\rmBS^{\rmS_2\rmI_1}) + K\calM^{\bar\rmB_k} \notag \\
& = 2P^{\rmS_2}(C^{\rmS_2}-C^{\rmS_1})(2M^2+4KM+8K) \notag \\
& \qquad + 2K(C^{\rmS_2}-C^{\rmS_1})(4L^2+2ML+3) + \frac 1 3KL(4L^2+15L-1).
\end{align}

The second type of input at the downlink $D_j$ (i.e., $\bdzeta_{D_j}^{\rmS_2\rmI_2}$) is generated by using \eqref{eq:r_tilde_S2}, \eqref{eq:r_bar_S2}, and \eqref{eq:D_jbar}.
The cost of \eqref{eq:r_tilde_S2} in each sub-frame is $\tilde\calA_{D_j}^{\rmS_2\rmI_2}=2P^{\rmS_2}M$ and $\tilde\calM_{D_j}^{\rmS_2\rmI_2}=4P^{\rmS_2}M$ real additions and multiplications, respectively.
The cost of \eqref{eq:r_bar_S2} in each sub-frame is $\bar\calA_{D_j}^{\rmS_2\rmI_2}=2M(2P^{\rmS_2}$ $-1)+\calA_{\rm pinv}^{(M,P^{\rmS_2})}$ real additions and $\bar\calM_{D_j}^{\rmS_2\rmI_2}=4MP^{\rmS_2}+\calM_{\rm pinv}^{(M,P^{\rmS_2})}$ real multiplications.
\textcolor{black}{The} LS estimator in \eqref{eq:D_jbar} has the same computational complexity as in \eqref{eq:B_k_bar}.
Hence, the number of real additions and multiplications, $\calA_{D_j}^{\rmS_2\rmI_2}$ and $\calM_{D_j}^{\rmS_2\rmI_2}$, required to generate the input $\bdzeta_{D_j}^{\rmS_2\rmI_2}$ are derived respectively by
\begin{align}
\calA_{D_j}^{\rmS_2\rmI_2}
& =(C^{\rmS_2}-C^{\rmS_1})(\tilde\calA_{D_j}^{\rmS_2\rmI_2} + \bar\calA_{D_j}^{\rmS_2\rmI_2}) + \calA^{\bar\rmD_j} \notag \\
& = \frac 2 3M(C^{\rmS_2}-C^{\rmS_1})(12MP^{\rmS_2}+3M^2+6P^{\rmS_2}-4) \notag \\
& \qquad + 2L(C^{\rmS_2}-C^{\rmS_1})(2M+4L-1) +\frac 2 3 L(3L^2-3M-1),
\end{align}
and
\begin{align}
\calM_{D_j}^{\rmS_2\rmI_2}
& =(C^{\rmS_2}-C^{\rmS_1})(\tilde\calM_{D_j}^{\rmS_2\rmI_2} + \bar\calM_{D_j}^{\rmS_2\rmI_2}) + \calM^{\bar\rmD_j} \notag \\
& = \frac 1 3M(C^{\rmS_2}-C^{\rmS_1})(24MP^{\rmS_2}+24P^{\rmS_2}+4M^2 \notag \\
& \qquad +15M-1) + 4L(C^{\rmS_2}-C^{\rmS_1})(M+2L)  +\frac 1 3 L(4L^2+15L-1).
\end{align}

\subsection{Complexity of ELM}
According to the proposed ELM-based \textcolor{black}{NN} framework in Section \ref{sec:DL-CE-appraoch}, the required \textcolor{black}{number} of real additions and multiplications, $\calA^{\rm ELM}$ and $\calM^{\rm ELM}$, for the ELM online testing is respectively  \textcolor{black}{expressed} as
\begin{align}
\calA^{\rm ELM} = N_\rmh(N_\zeta+N_\gamma+1)-N_\gamma,
\end{align}
and
\begin{align}
\calM^{\rm ELM} = N_\rmh(N_\zeta+N_\gamma).
\end{align}

\section{Simulation Results}\label{sec:simulation}
In this section, extensive simulations are provided to assess the performance of the proposed ELM-based channel estimation approach.
The LS estimator is employed as the benchmark scheme for comparison.
Firstly, the simulation setup for the IRS-assisted multi-user ISAC system is introduced.
Then, the proposed ELM-based approach is evaluated under \textcolor{black}{various} SNR conditions and system parameters to unveil its estimation/generalization performance.
Finally, the computational complexity of the proposed ELM-based approach is quantitatively measured in terms of various system parameters.

\subsection{Simulation Setup}
For the \textcolor{black}{narrowband} IRS-assisted multi-user ISAC system, consider $M=6$, $L=30$, $K=J=6$, $C^{\rmS_1}=1$, and $C^{\rmS_2}=L+1$ unless further specified.
{This system operates at the $3.5\,\rm GHz$ frequency band, which is also used for 5G new radio and has the profound potential to be applied in the ISAC systems \cite{ref:major-3.5G}.}
\textcolor{black}{The time slot duration (i.e., pilot symbol duration) is set to $T_{\rm P}=0.52\,\rm \mu s$ according to the 3-rd generation partnership project standard \cite{ref:major-3gpp.36.211}.}
Since each sub-frame contains $P^{\rmS_\ell}$ time slots, the sub-frame duration is devised as $T_{\rm F}^{\rmS_\ell}=T_{\rm P}P^{\rmS_\ell}$ in the estimation stage $\rmS_\ell$.
With $P^{\rmS_1}=M+K$ and $P^{\rmS_2}=\max\{M,K\}$, the sub-frame duration in $\rmS_1$ and $\rmS_2$ is respectively obtained as $T_{\rm F}^{\rmS_1}=T_{\rm P}P^{\rmS_1}= 6.24\,\rm \mu s$ and $T_{\rm F}^{\rmS_2}=T_{\rm P}P^{\rmS_2} = 3.12\,\rm \mu s$.
Correspondingly, the total sub-frame duration required for the two estimation stages is $T_{\rm E}=C^{\rmS_1}T_{\rm F}^{\rmS_1} + (C^{\rmS_2}-C^{\rmS_1})T_{\rm F}^{\rmS_2}= 99.84\,\rm \mu s$.
As in \cite{ref:major-coherent-time}, the channel coherence time is set to $T_{\rm coh} = 1\,\rm ms$.
It is obvious that $T_{\rm E}$ is much smaller than $T_{\rm coh}$ (i.e., $99.84\,\rm \mu s \ll 1\,\rm ms$).
Thus, the SAC channels are unchanged during the two estimation stages.
Based on the radar channel model in \cite{ref:ChModel-b}, the sensing channel $\bfA$ is modeled as
\begin{align}
\bfA = \alpha_\rmS \bfa(\vartheta_\rmS)\bfa(\vartheta_\rmS)^\rmT\textcolor{black}{.}
\label{eq:A_model}
\end{align}
\textcolor{black}{
Since the narrowband transmission is considered and the pilot symbol duration of the sensing signal is much smaller than the channel coherence time (i.e., $T_{\rm P} \ll T_{\rm coh}$), the amplitude of $\bfA$ is assumed to be approximately constant during the offline training and online testing phases \cite{ref:ChModel-b,ref:channel-b-fixed-amplitude}. 
On the other hand, the phase-shift of $\bfA$ is sensitive to the propagation delay, and a small variation in propagation delay may cause a large phase difference.
Due to the above reasons, the complex-valued radar cross $\alpha_\rmS$ in $\bfA$ is with constant amplitude of $|\alpha_\rmS|=1$, whereas its phase-shift is uniformly distributed from $[0,2\pi)$ \cite{ref:ChModel-b,ref:channel-b-fixed-amplitude}.}
Regarding the $\bar M$ array elements, the steering vector $\bfa(\bar\vartheta)$ associated with the azimuth angle $\bar\vartheta$ is formulated as
\begin{align}
\bfa(\bar\vartheta) = [1, e^{\jmath\frac{2\pi {\bar d}}{\lambda}\sin(\bar\vartheta)}, \ldots, e^{\jmath\frac{2\pi {\bar d}}{\lambda}({\bar M}-1)\sin(\bar\vartheta)}]^\rmT,
\label{eq:steering_vec}
\end{align}
where ${\bar d}$ and $\lambda$ denote the inter-element spacing and signal wavelength, respectively.
As such, $\bfa(\vartheta_\rmS)$ in \eqref{eq:A_model} is obtained according to \eqref{eq:steering_vec} by letting $\bar\vartheta=\vartheta_\rmS$, $\bar d=d_\rmB$, and $\bar M=M$.
Here, $\vartheta_\rmS$ and $d_\rmB$ are the azimuth angle of the target and antenna spacing of the ISAC BS, respectively.
\textcolor{black}{The communication channels are modeled as Rician fading, referring to \cite{ref:ChModel-refpower} and \cite{ref:DL-IRS-ChE-TWC}.}
Therefore, the IRS-BS channel with \textcolor{black}{the} Rician factor $K_\rmIB=10$ is \textcolor{black}{described} by
\begin{align}
\bfH = \sqrt{\frac {K_\rmIB}{K_\rmIB+1}} \bfH_\rmLoS
+ \sqrt{\frac {1}{K_\rmIB+1}} \bfH_\rmNLoS,
\label{eq:H_rician}
\end{align}
where $\bfH_\rmLoS$ and $\bfH_\rmNLoS$ respectively represent the line-of-sight (LoS) and non-LoS (NLoS) components of $\bfH$. As in \cite{ref:ChModel-refpower}, define $\bfH_\rmLoS=\bfa(\vartheta_\rmB)\bfa(\vartheta_\rmI)^\rmH$ and let the entries in $\bfH_\rmNLoS$ follow ${\cal CN}(0,1)$.
According to the steering vector formulated in \eqref{eq:steering_vec}, $\bfa(\vartheta_\rmB)$ is derived by setting $\bar\vartheta=\vartheta_\rmB$, $\bar d=d_\rmB$, and $\bar M=M$, whereas $\bfa(\vartheta_\rmI)$ is obtained with $\bar\vartheta=\vartheta_\rmI$, $\bar d=d_\rmI$, and $\bar M=L$.
Here, $\vartheta_\rmB$ and $\vartheta_\rmI$ respectively denote the angle of arrival (AoA) and angle of departure (AoD) related to the IRS-BS channel.
Without loss of generality, the BS antenna spacing $d_\rmB$ and IRS inter-element spacing $d_\rmI$ satisfy $d_\rmB=d_\rmI=\frac \lambda 2$.
The rest of the communication channels are generated similarly as in \eqref{eq:H_rician}, and their Rician factors are set to $10$.
Particularly, the LoS components of the uplink channels (i.e., $\bfb_k$ and $\bfg_k$) are respectively given by  $\bfb_{k,\rmLoS}=\bfa(\vartheta_{U_k\rmB})$ with AoA $\vartheta_{U_k\rmB}$ and $\bfg_{k,\rmLoS}=\bfa(\vartheta_{U_k\rmI})$ with AoA $\vartheta_{U_k\rmI}$, while that of the downlink channels (i.e., $\bfd_j$ and $\bff_j$) are respectively denoted by $\bfd_{j,\rmLoS}=\bfa(\vartheta_{\rmB D_j})$ with AoD $\vartheta_{\rmB D_j}$ and $\bff_{j,\rmLoS}=\bfa(\vartheta_{\rmI D_j})$ with AoD $\vartheta_{\rmI D_j}$.
Furthermore, to describe the path loss factor of each channel, a distance-dependent path loss model is employed.
For the sensing channel, the path loss at distance $d_\rmS$ is modeled as $\xi_\rmS = \xi_0(\frac{d_\rmS}{d_0})^{-\beta_\rmS}$, where $\xi_0=-30\,\rm dBm$ represents the path loss at the reference distance $d_0=1\,\rm m$. 
The path losses of the communication channels (i.e., $\xi_{\rmI\rmB}$, $\xi_{U_k\rmB}$, $\xi_{U_k\rmI}$, $\xi_{\rmB D_j}$, and $\xi_{\rmI D_j}$) are formulated similarly as the sensing one.
The distances of BS-target-BS, IRS-BS, $U_k$-BS, BS-$D_j$, $U_k$-IRS, IRS-$D_j$ are set to $d_\rmS=150\,\rm m$, $d_{\rmIB}=d_{U_k\rmB}=d_{\rmB D_j}=50\,\rm m$, and $d_{U_k\rmI}=d_{\rmI D_j}=2\,\rm m$, respectively.
The path loss exponents are $\beta_\rmS=3$, $\beta_{\rmIB}=2.3$, $\beta_{U_k\rmB}=\beta_{\rmB D_j}=3.5$, $\beta_{U_k\rmI}=\beta_{\rmI D_j}=2$ \cite{ref:ChModel-refpower}.
Correspondingly, the AoA/AoD angles are $\vartheta_\rmS=-\pi$, $\vartheta_\rmI=\vartheta_\rmB=-\frac \pi 2$, $\vartheta_{U_k\rmB}$ $=\arcsin(\frac {d_\rmIB} {2d_{U_k\rmB}})-\pi$, $\vartheta_{U_k\rmI}$ $=\pi-\arcsin(\frac {d_\rmIB} {2d_{U_k\rmI}})$, $\vartheta_{\rmB D_j}$ $=\arcsin(\frac {d_\rmIB} {2d_{\rmB D_j}})$, and $\vartheta_{\rmI D_j}$ $=-\arcsin(\frac {d_\rmIB} {2d_{\rmI D_j}})$.
\textcolor{black}{The propagation delay of the sensing signal is derived by $T_{\rmS}=\frac{d_\rmS}{v_{\rm c}}=0.5\,\rm\mu s$, where $v_{\rm c}=3 \times {10^8} \,\rm m/s$ is the speed of the electromagnetic wave.
Similarly, the propagation delays of the uplink and downlink communication signals are respectively obtained as $T_{U_k}=T_{D_j}\approx0.17\,\rm\mu s$.
Since the pilot symbol duration $T_{\rm P}=0.52\,\rm \mu s$ is larger than $T_{\rmS}$, $T_{U_k}$, and $T_{D_j}$, the received SAC signals at the ISAC BS and downlink $D_j$ are synchronized, as formulated in \eqref{eq:y_cp_Orig} and \eqref{eq:r_jcp_Orig}.}
As in \cite{ref:IRS-ISAC-ChEst-TVT,ref:ChModel-power}, the transmit power of the ISAC BS and uplink UEs is set to $\mathcal P_\rmB=20\,\rm dBm$ and $\mathcal P_\rmU=15\,\rm dBm$, respectively.


Table \ref{table:ELM} summarizes the hyperparameters of the proposed DE-ELM and RE-ELM.
To train the proposed ELM networks, the training dataset contains $N_{\rm tr}=VQ=10^4$ samples for each SNR condition with $V=10^3$, $Q=10$, and $\text{SNR}_\text{ch}=30\,\rm dB$.
For the online testing phase, different $N_{\rm te}=10^3$ samples are adopted to assess the estimation and generalization performance for various SNR conditions.
Note that the SNR in the simulations refers to that at the receiver side.
Thus, the SNR at the ISAC BS and downlink $D_j$ is respectively denoted by $\text{SNR}_\rmB = \frac{{\cal R}_\rmB}{\sigma^2}$ and $\text{SNR}_{D_j} = \frac{{\cal R}_{D_j}}{\varsigma^2}$, where ${\cal R}_\rmB$ and ${\cal R}_{D_j}$ are the received signal power.
For the two estimation stages (i.e., $\rmS_1$ and $\rmS_2$), ${\cal R}_\rmB$ is respectively equal to ${\cal P}_\rmB\xi_\rmS+{\cal P}_\rmU\xi_{U_k\rmB}$ and ${\cal P}_\rmB\xi_\rmS+{\cal P}_\rmU(\xi_{U_k\rmB}+\xi_{U_k\rmI}\xi_{\rmI\rmB})$, while ${\cal R}_{D_j}$ are respectively given by ${\cal P}_\rmB\xi_{\rmB D_j}$ and ${\cal P}_\rmB(\xi_{\rmB D_j}+\xi_{\rmI D_j}\xi_{\rmI\rmB})$.
Furthermore, the normalized mean square error (NMSE) is employed as an estimation performance metric, and is denoted by
\begin{align}
\text{NMSE}=\mathbb{E}\bigg\{ \frac{\| \text{Estimated} - \text{Actual} \|_F^2}{\| \text{Actual} \|_F^2} \bigg\}.
\label{eq:NMSE}
\end{align}



\subsection{NMSE versus SNR}

{In \textcolor{black}{Fig. \ref{fig:NMSE_SNR_BS}}, the NMSE performance of the proposed {ELM-based} approach is compared with the LS and two \textcolor{black}{NN}-based benchmark schemes (i.e., CNN-based and FNN-based) under different SNR values.
The hyperparameters used in the CNN-based benchmark scheme refer to the state-of-the-art work in \cite{ref:IRS-ISAC-ChEst-TVT}.
\textcolor{black}{The FNN-based benchmark scheme employs the same optimizer and stopping criterion as in \cite{ref:IRS-ISAC-ChEst-TVT} to train the FNN, while its hidden layer size and activation functions are the same as the proposed ELM network.}
Moreover, Table \ref{table:NMSE_SNR} compares the training time of the proposed ELM-based approach with two \textcolor{black}{NN}-based benchmark schemes. 
All the NNs are realized and trained on a computer with an i7-7500U 2.90 GHz central processing unit under the same operation conditions.} 
The SNR region in the offline training phase is considered as $\text{SNR} = [15,20]\,\rm{dB}$ with a step size of $5\,\rm{dB}$, while that in the online testing phase is set to $\text{SNR} = [-10,20]\,\rm{dB}$ with a step size of $2.5\,\rm{dB}$.

\begin{figure}
\centering
\subfigure[ ]
{\begin{minipage}[b]{0.45\textwidth}
\includegraphics[width=1\textwidth]{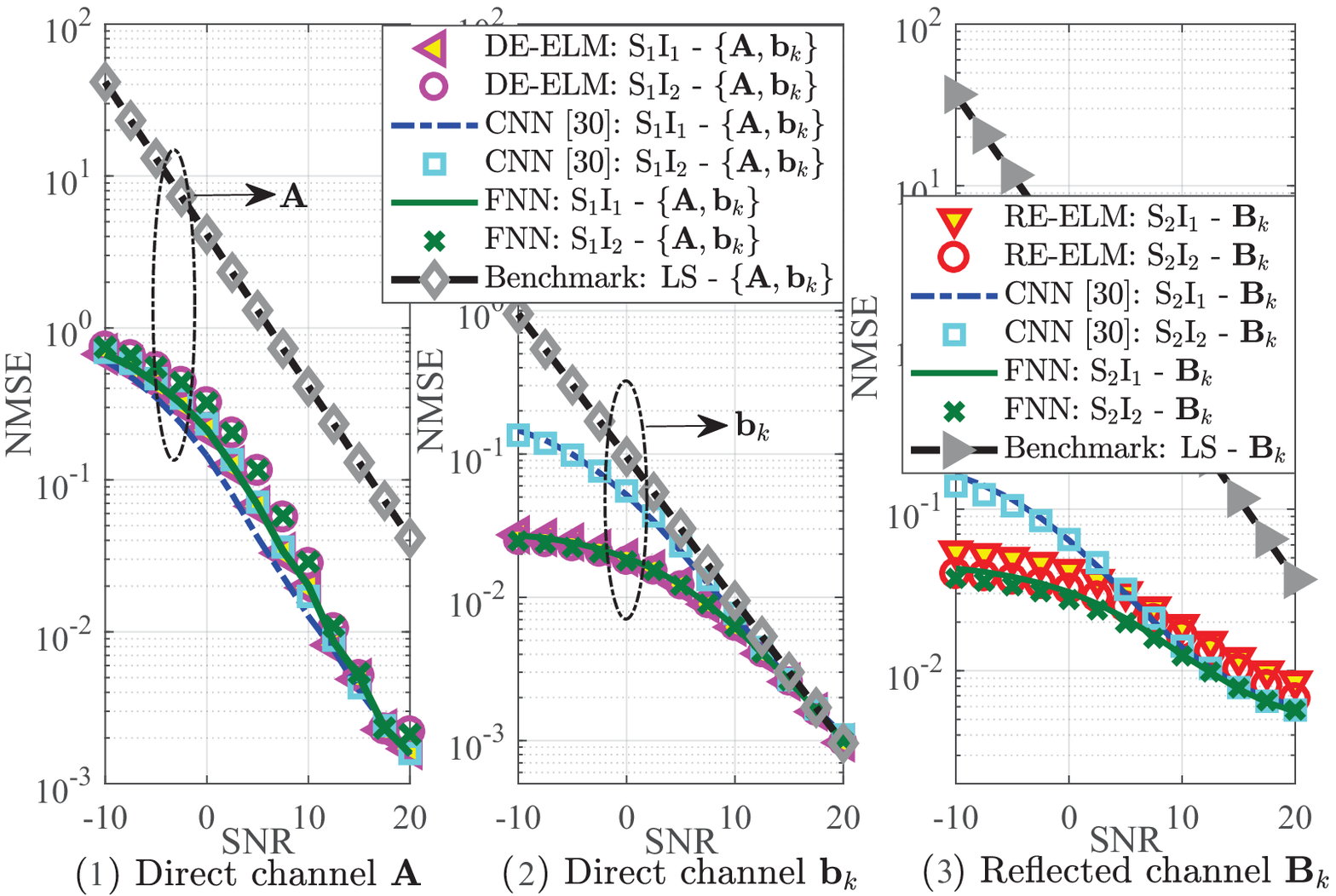}
\end{minipage}}
\subfigure[ ]
{\begin{minipage}[b]{0.45\textwidth}
\includegraphics[width=1\textwidth]{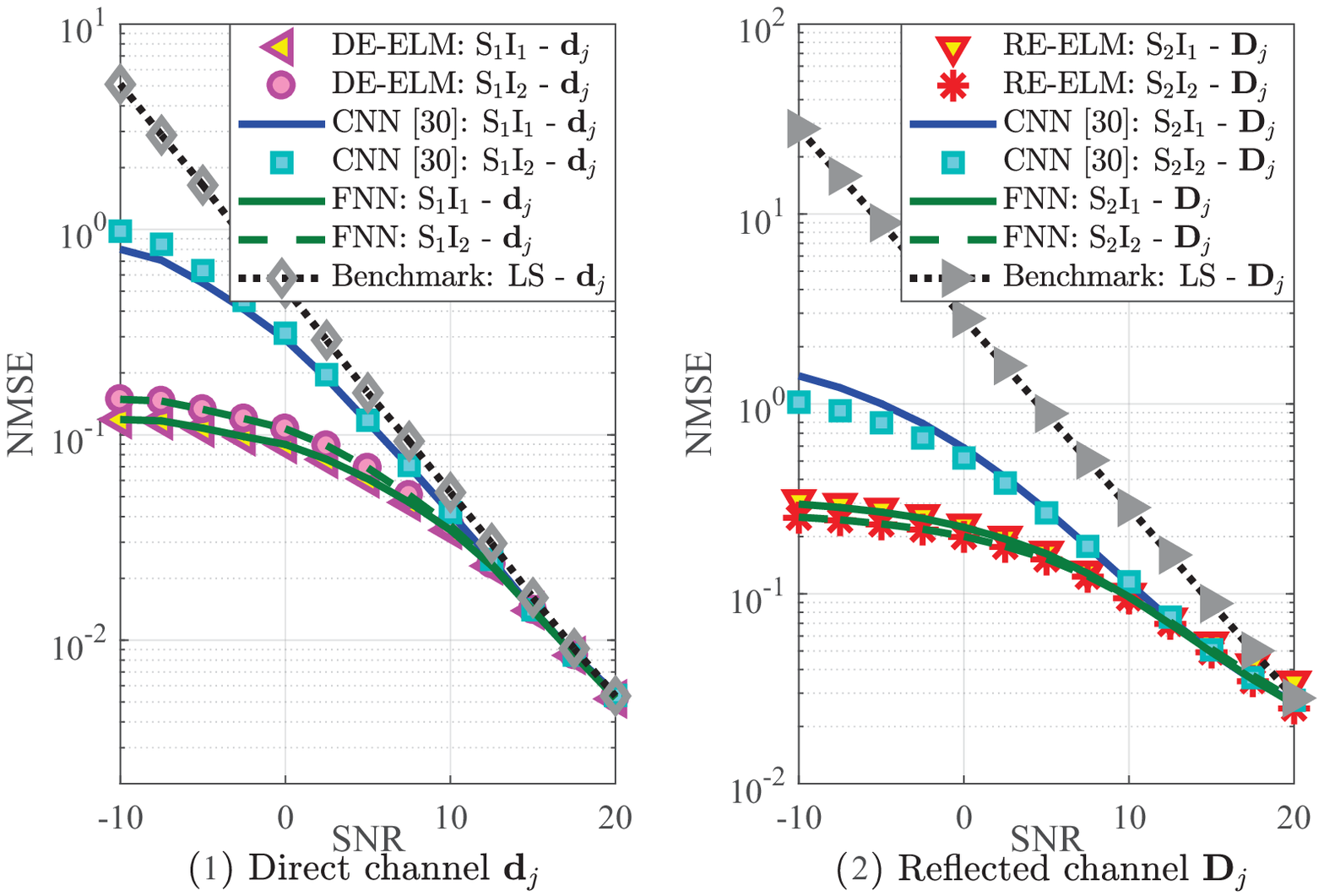}
\end{minipage}}
\caption{\textcolor{black}{NMSE of SAC channels estimation versus SNR for $M=6$ and $L=30$: (a) SAC channels at the ISAC BS, (b) Communication channels at the downlink $D_j$.}}
\label{fig:NMSE_SNR_BS}
\end{figure}

\textcolor{black}{Fig. \ref{fig:NMSE_SNR_BS}(a)} investigates the estimation performance for the SAC channels (i.e., $\bfA$, $\bfb_k$, and $\bfB_k$) at the ISAC BS.
As can be observed, the NMSE performance of the proposed DE-ELM and RE-ELM outperforms the LS benchmark scheme.
In addition, when trained by the different types of input-output pairs, the proposed \textcolor{black}{ELM-based} approach provides comparable estimation performance to the LS benchmark scheme.
\textcolor{black}{It is noted that for estimating $\bfA$, $\bfb_k$, and $\bfB_k$, the NMSE improvement achieved by the proposed {ELM-based} approach is around $10^{1.5}\,\rm x$, $10^{1.9}\,\rm x$, and $10^{2.7}\,\rm x$, respectively, compared to the LS benchmark scheme at $\text{SNR}=-10\,\rm dB$}. 
\textcolor{black}{Additionally, the proposed ELM-based approach obtains significant NMSE performance improvement compared to the CNN-based benchmark scheme when estimating $\bfb_k$ and $\bfB_k$, and their NMSE performance is comparable for estimating $\bfA$.
The above findings indicate that the learning capacity of the proposed ELM-based is considerable, even compared with the more complex CNN-based benchmark scheme in \cite{ref:IRS-ISAC-ChEst-TVT}.}
\textcolor{black}{The corresponding training time of the proposed ELM-based approach is less than $15$ seconds and extremely faster \textcolor{black}{when compared with} that of the CNN-based benchmark schemes, as seen from Table \ref{table:NMSE_SNR}.
In addition, the online testing computational complexity of the proposed ELM-based approach is lower than \textcolor{black}{that of} the CNN-based benchmark scheme due to its simpler NN structure.
}
{By taking another look at \textcolor{black}{Fig. \ref{fig:NMSE_SNR_BS}(a)} and Table \ref{table:NMSE_SNR}, the NMSE performance of the proposed ELM-based approach is comparable to \textcolor{black}{that of} the FNN-based benchmark scheme for all SAC channels (i.e., $\bfA$, $\bfb_k$, and $\bfB_k$), while the advantages of the proposed ELM on the training time are quite apparent.
The reason is that compared to the FNN-based benchmark scheme with the same NN structure, the proposed ELM-based approach learns the weights and biases based on an inverse operation instead of the gradient-based optimizers \textcolor{black}{as} in \textcolor{black}{the} FNN-based benchmark, promoting the training speed.
As for the online testing computational complexity, the proposed ELM-based approach performs the same as the FNN-based benchmark scheme due to their identical NN size (i.e., tensor size of the input, output, and hidden layers, as well as activation functions).
In summary, at the same NN complexity and estimation accuracy, the ELM-based approach significantly reduces the training (learning) time compared to the FNN-based benchmark. 



\textcolor{black}{Fig. \ref{fig:NMSE_SNR_BS}(b)} evaluates the estimation performance for the downlink communication channels (i.e., $\bfd_j$ and $\bfD_j$) at the downlink $D_j$.
As can be observed, the proposed {ELM-based} approach trained by the first type of input-output pair achieves comparable NMSE performance as that with the second one, whereas both of them are superior to the LS benchmark scheme.
\textcolor{black}{For instance, compared with the LS benchmark scheme at $\text{SNR}=-10\,\rm dB$, the proposed DE-ELM and RE-ELM respectively attain $10^{1.4}\,\rm x$ and $10^{2}\,\rm x$ NMSE improvements.} 
{For the estimation of both $\bfd_j$ and $\bfD_j$ in \textcolor{black}{Fig. \ref{fig:NMSE_SNR_BS}(b)}, the NMSE performance of the proposed ELM-based approach outperforms the CNN-based benchmark scheme, while it is comparable to the FNN-based one.} 
\textcolor{black}{The corresponding training time of the proposed ELM-based approach \textcolor{black}{for} downlink $D_j$ is significantly faster than the CNN-based and FNN-based benchmark schemes, as depicted in Table \ref{table:NMSE_SNR}. 
Regarding the online testing computational complexity, \textcolor{black}{the one for} the proposed ELM-based approach is lower than \textcolor{black}{for} the complex CNN-based benchmark scheme and the same as the FNN-based one.
The findings \textcolor{black}{presented} in \textcolor{black}{Fig. \ref{fig:NMSE_SNR_BS}} and Table \ref{table:NMSE_SNR} unveil the superiorities of the proposed ELM-based approach in terms of the NN training, NN testing, and estimation performance, enabling it to be a low-cost channel estimator at both ISAC BS and downlink UEs.}


\begin{table}[t] 
\newcommand{\tabincell}[2]{\begin{tabular}{@{}#1@{}}#2\end{tabular}}
\centering
\caption{{Training Time Comparison of the Proposed ELM-based Approach and {NN}-based Benchmark Schemes.}} \label{table:NMSE_SNR}
\begin{tabular}{p{1.7cm}<{\centering}| p{2.5cm}<{\centering}| p{2.2cm}<{\centering}| p{2.2cm}<{\centering} | p{2.2cm}<{\centering} }
\hline
  \multicolumn{2}{c|}{\bf Training time (in seconds)} & {\bf ELM-based} & {\bf CNN-based \cite{ref:IRS-ISAC-ChEst-TVT}} & {\bf FNN-based} \\
\hline
\hline
\multirow{4}{*}{\bf \tabincell{c}{ISAC BS\\(Fig. \ref{fig:NMSE_SNR_BS})}} & $\rmS_1\rmI_1$ - $\{\bfA,\bfb_k\}$ & 2.36 & 95.13 & 10.26 \\ \cline{2-5}
                              & $\rmS_1\rmI_2$ - $\{\bfA,\bfb_k\}$ & 3.16  & 129.47   & 12.80 \\ \cline{2-5}
                              & $\rmS_2\rmI_1$ - $\bfB_k$          & 14.65 & 23253.44 & 485.37 \\ \cline{2-5}
                              & $\rmS_2\rmI_2$ - $\bfB_k$          & 14.53 & 25216.30 & 357.01 \\ \cline{1-5}
\multirow{4}{*}{\bf \tabincell{c}{Downlink $D_j$\\(Fig. \ref{fig:NMSE_SNR_UE})}} & $\rmS_1\rmI_1$ - $\bfd_j$ & 2.35 & 28.78 & 3.81 \\ \cline{2-5}
                              & $\rmS_1\rmI_2$ - $\bfd_j$          & 2.12 & 7.17    & 2.27 \\ \cline{2-5}
                              & $\rmS_2\rmI_1$ - $\bfD_j$          & 9.34 & 1488.38 & 88.81 \\ \cline{2-5}
                              & $\rmS_2\rmI_2$ - $\bfD_j$          & 6.55 & 1201.15 & 67.34 \\ \cline{2-5}
\hline
\end{tabular}
\end{table} 

\subsection{NMSE versus Channel Dimension}

\subsubsection{NMSE versus L}\label{sec:NMSE_L}
Since $L$ is an essential parameter that affects the reflected channel dimension, \textcolor{black}{Fig. \ref{fig:NMSE_L_BS} illustrates} the impact of varying $L$ on the NMSE performance.
The SNR conditions in the offline training and online testing phases are fixed to $\text{SNR} = 0\,\rm{dB}$ and $\text{SNR} = 10\,\rm{dB}$, respectively, for the proposed RE-ELM.

\textcolor{black}{Fig. \ref{fig:NMSE_L_BS}(a)} shows the estimation performance of the uplink reflected channel $\bfB_k$ at the ISAC BS.
Obviously, for different $L$ values and SNR conditions, the proposed RE-ELM significantly enhances the estimation accuracy of $\bfB_k$ compared with the LS benchmark scheme.
It is also noted that the NMSE performance of the proposed RE-ELM slightly increases as $L$ increases.
This lies in that with the increase of channel dimension, the mapping of the input-output pairs (i.e., $(\bdzeta_\rmBS^{\rmS_2\rmI_1},\bdgamma_\rmBS^{\rmS_2})$ and $(\bdzeta_\rmBS^{\rmS_2\rmI_2},\bdgamma_\rmBS^{\rmS_2})$) is more challenging to be learned precisely by the proposed RE-ELM. 

\textcolor{black}{Fig. \ref{fig:NMSE_L_BS}(b)} depicts the NMSE performance of the downlink reflected channel $\bfD_j$ at the downlink $D_j$.
As can be observed, under different $L$ and SNR setups, the proposed RE-ELM outperforms the LS benchmark scheme for estimating $\bfD_j$.
Moreover, the proposed RE-ELM trained by the two types of input-output pairs (i.e., $(\bdzeta_{D_j}^{\rmS_2\rmI_1},\bdgamma_{D_j}^{\rmS_2})$ and $(\bdzeta_{D_j}^{\rmS_2\rmI_2},\bdgamma_{D_j}^{\rmS_2})$) is robust to the change of $L$.



\begin{figure}
\centering
\subfigure[ ]
{\begin{minipage}[b]{0.45\textwidth}
\includegraphics[width=1\textwidth]{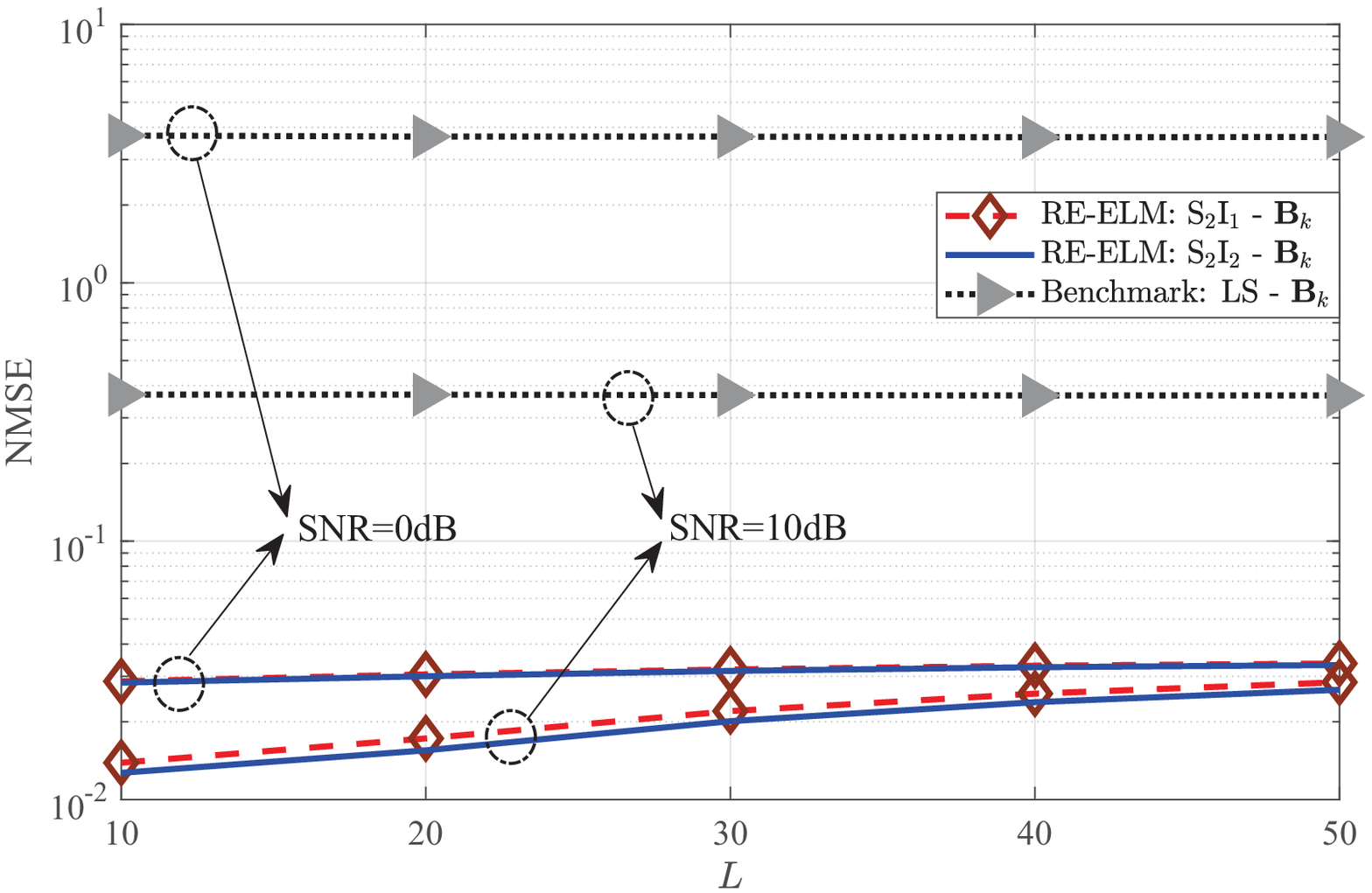}
\end{minipage}}
\subfigure[ ]
{\begin{minipage}[b]{0.45\textwidth}
\includegraphics[width=1\textwidth]{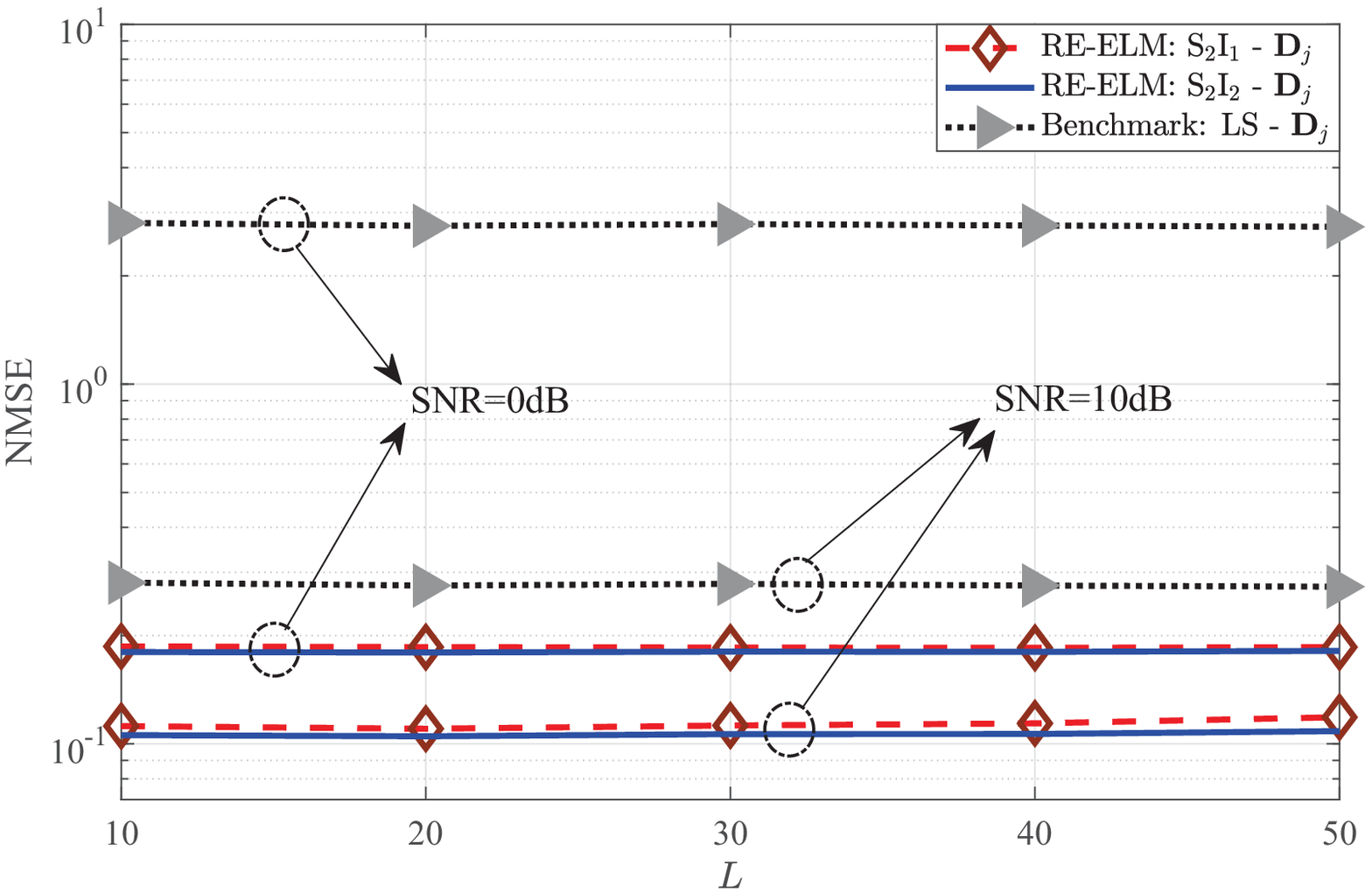}
\end{minipage}}
\caption{\textcolor{black}{NMSE of communication channels estimation versus $L$ for $M=6$ under $\text{SNR}=0\,\rm dB$ and $10\,\rm dB$: (a) Uplink reflected communication channel at the ISAC BS, (b) Downlink reflected communication channel at the downlink $D_j$.}}
\label{fig:NMSE_L_BS}
\end{figure}

\subsubsection{NMSE versus M}

\textcolor{black}{Fig. \ref{fig:NMSE_M_BS}} investigate the effect of increasing $M$ on the estimation performance.
The SNR setups are \textcolor{black}{similar to} Section \ref{sec:NMSE_L} with $L=15$.
As seen from \textcolor{black}{Fig. \ref{fig:NMSE_M_BS}(a)}, the proposed {ELM-based} approach at the ISAC BS is superior to the LS benchmark scheme when estimating the SAC channels (i.e., $\bfA$, $\bfb_k$, and $\bfB_k$) under different $M$ and SNR values.
The NMSE of $\bfA$ achieved by the proposed DE-ELM decreases as $M$ increases since the DE-ELM can extract more distinguishable features \textcolor{black}{for} a larger $M$.
Similar NMSE improvement can be observed when estimating $\bfb_k$. 
For the uplink reflected channel estimation of $\bfB_k$ in \textcolor{black}{Fig. \ref{fig:NMSE_M_BS}(a)}, when $M$ increases, the proposed RE-ELM trained by the second type of input-output pair (i.e., $(\bdzeta_\rmBS^{\rmS_2\rmI_2},\bdgamma_\rmBS^{\rmS_2})$) provides {an enhanced} NMSE performance {when compared with} the first one (i.e., $(\bdzeta_\rmBS^{\rmS_2\rmI_1},\bdgamma_\rmBS^{\rmS_2})$).
The reason is that $\bdzeta_\rmBS^{\rmS_2\rmI_2}$ in \eqref{eq:zeta_S2I2_BS} is generated by the LS estimation results (i.e., $\bar\bfB_k$ in \eqref{eq:B_k_bar}), whereas $\bdzeta_\rmBS^{\rmS_2\rmI_1}$ in \eqref{eq:zeta_S2I1_BS} relies on the received signals without signal pre-processing (i.e., $\bfY_c^{\rmS_2}$ in \eqref{eq:Y_c_S2}).
In other words, the mapping of the second type of input-output pair is easier than \textcolor{black}{for} the first one.
Therefore, the proposed RE-ELM with $(\bdzeta_\rmBS^{\rmS_2\rmI_2},\bdgamma_\rmBS^{\rmS_2})$ obtains a better feature extraction ability, \textcolor{black}{which enhances} the estimation performance.



\begin{figure}
\centering
\subfigure[ ]
{\begin{minipage}[b]{0.45\textwidth}
\includegraphics[width=1\textwidth]{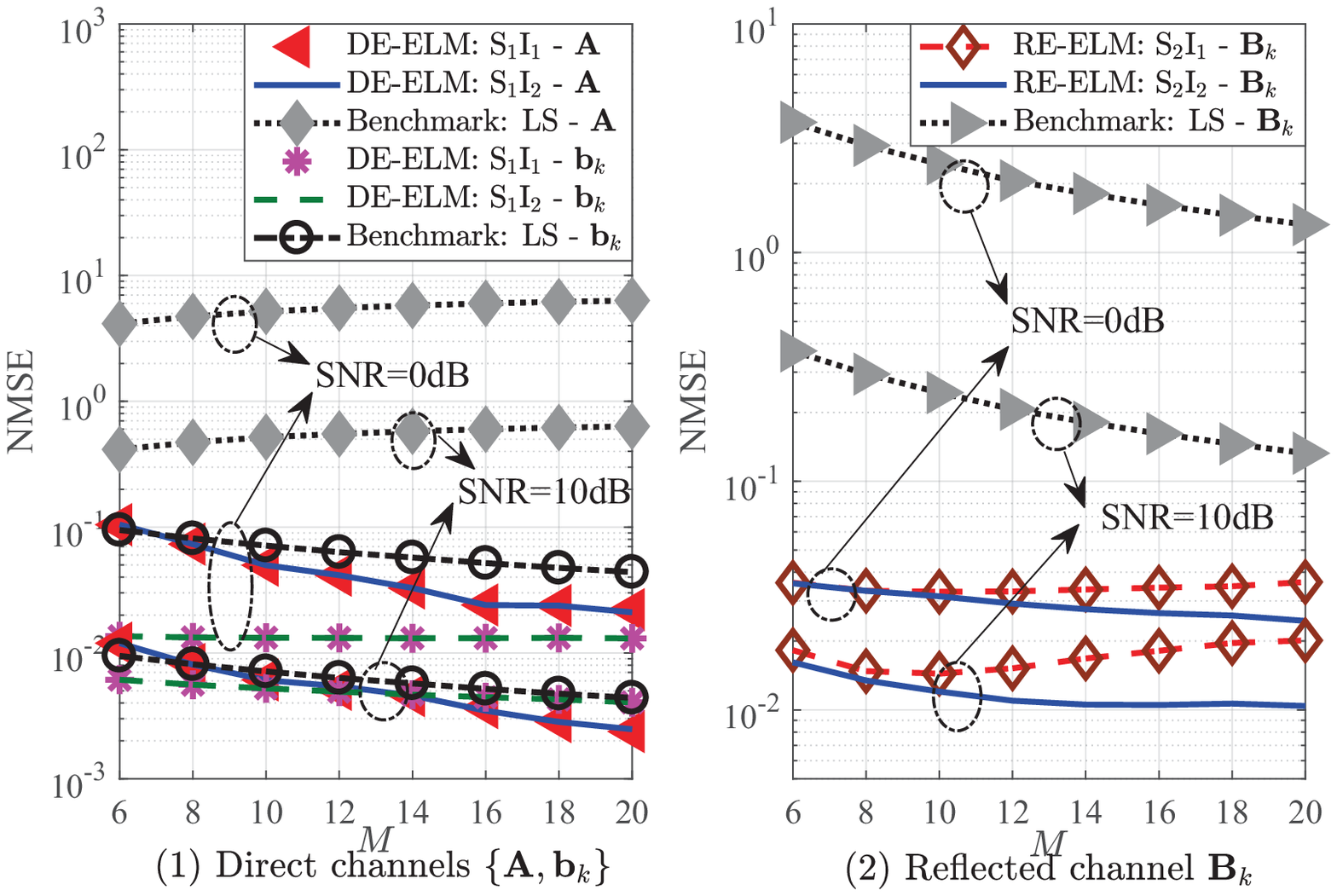}
\end{minipage}}
\subfigure[ ]
{\begin{minipage}[b]{0.45\textwidth}
\includegraphics[width=1\textwidth]{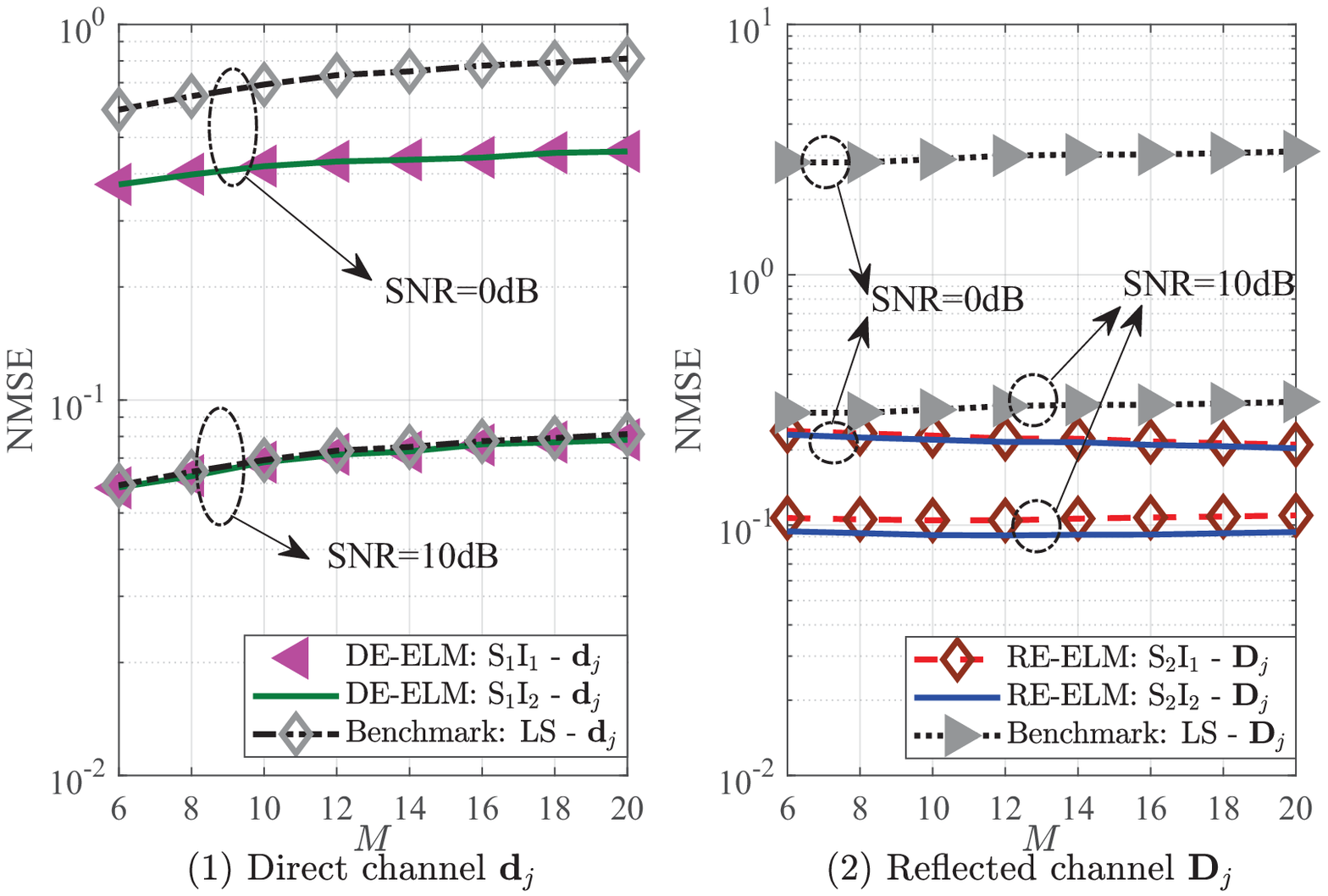}
\end{minipage}}
\caption{\textcolor{black}{NMSE of SAC channels estimation versus $M$ for $L=15$ under $\text{SNR}=0\,\rm dB$ and $10\,\rm dB$: (a) SAC channels at the ISAC BS, (b) Communication channels at the downlink $D_j$.}}
\label{fig:NMSE_M_BS}
\end{figure}


\textcolor{black}{Fig. \ref{fig:NMSE_M_BS}(b)} depicts the estimation performance of the direct and reflected downlink channels (i.e., $\bfd_j$ and $\bfD_j$) at the downlink $D_j$.
Obviously, the proposed \textcolor{black}{ELM-based} approach achieves comparable NMSE performance to the LS benchmark scheme when adopting the two types of input-output pairs, while it outperforms the LS benchmark scheme. 
For the estimation of $\bfd_j$, the NMSE of the proposed DE-ELM slightly increases as $M$ increases.
This can be attributed to two aspects, as follows:
First, the increase of the channel dimension makes the proposed DE-ELM harder to extract the efficient features of $\bfd_j$.
Second, the downlink $D_j$ estimates the channels individually (i.e., without cooperative estimation gain).
The above reasons affect the estimation accuracy of the proposed DE-ELM at the downlink $D_j$.
For the estimation of $\bfD_j$, the proposed RE-ELM is insensitive to the increase of $M$, consistent with the finding in \textcolor{black}{Fig. \ref{fig:NMSE_L_BS}(b)}.
The results from \textcolor{black}{Figs. \ref{fig:NMSE_M_BS}(b) and \ref{fig:NMSE_L_BS}(b)} indicate that the proposed RE-ELM at the downlink $D_j$ possesses substantial scalability under a wide range of channel dimensions (i.e., $L$ and $M$).


\subsection{Computational Complexity Assessment}

\begin{figure}
\centering
\subfigure[ ]
{\begin{minipage}[b]{0.45\textwidth}
\includegraphics[width=1\textwidth]{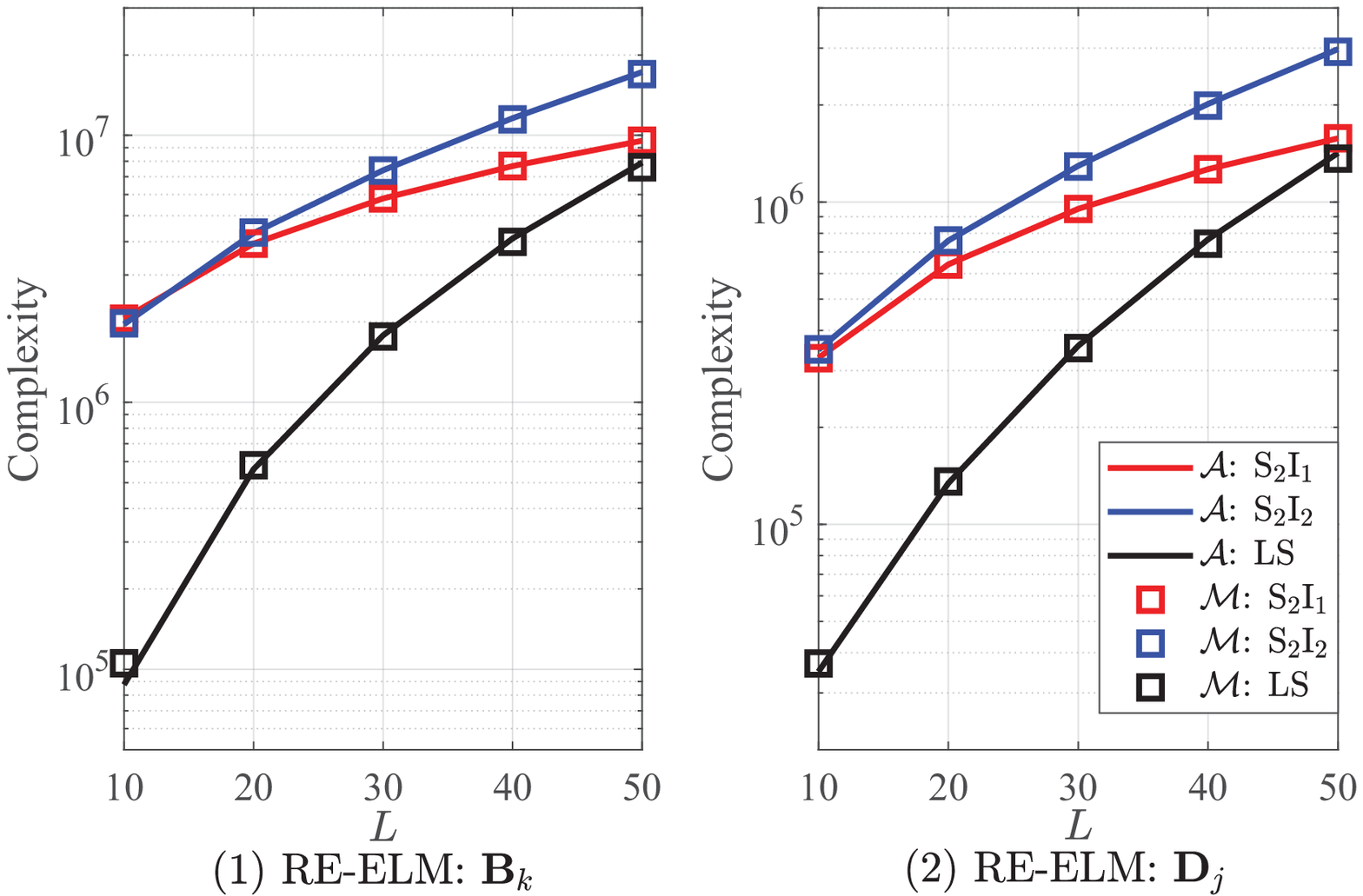}
\end{minipage}}
\subfigure[ ]
{\begin{minipage}[b]{0.45\textwidth}
\includegraphics[width=1\textwidth]{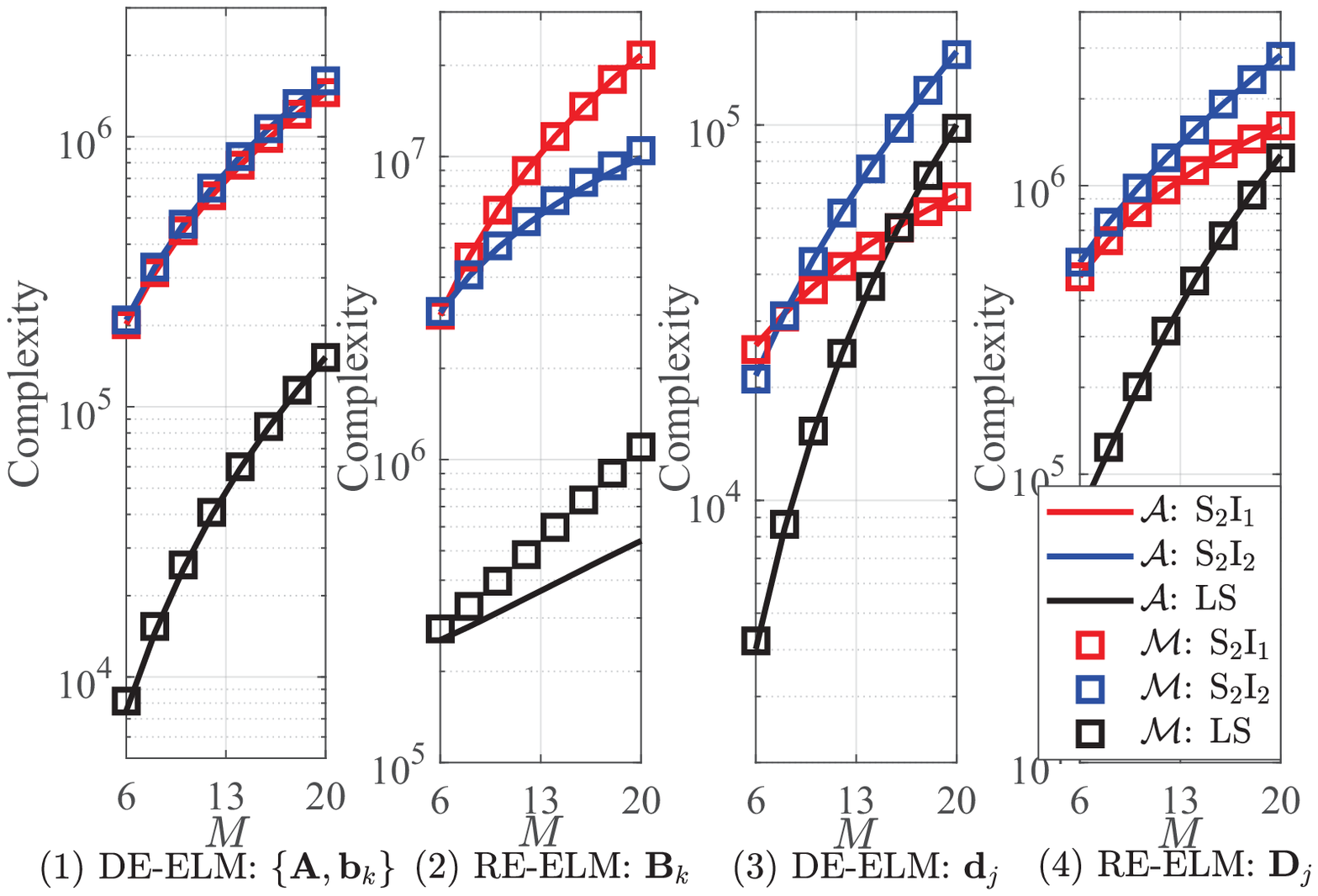}
\end{minipage}}
\caption{\textcolor{black}{Complexity comparison for different channel dimension: (a) Complexity versus $L$ for $M=6$, (b) Complexity versus $M$ for $L=15$.}}
\label{fig:Complexity_BS_UE}
\end{figure}

In \textcolor{black}{Fig. \ref{fig:Complexity_BS_UE}}, the computational complexity of the proposed {ELM-based} approach is compared to the LS benchmark scheme in terms of the required number of real additions and multiplications (i.e., $\cal A$ and $\cal M$), respectively.
According to the {closed-form expressions derived} in Section \ref{sec:complexity}, the costs of the inputs generation and ELM online testing are summed up to evaluate the complexity of the proposed ELM-based approach.

\textcolor{black}{Fig. \ref{fig:Complexity_BS_UE}(a)} assesses the complexity for estimating the reflected channels (i.e., $\bfB_k$ and $\bfD_j$) when increasing $L$, while \textcolor{black}{Fig. \ref{fig:Complexity_BS_UE}(b)} shows the complexity for estimating all the SAC channels (i.e., $\bfA$, $\bfb_k$, $\bfB_k$, $\bfd_j$, and $\bfD_j$) as $M$ increases.
As seen from these figures, the {computational} complexity of the proposed {ELM-based} approach is comparable to the LS benchmark scheme when the channel dimension becomes larger.
Moreover, for estimating $\bfB_k$ under different $M$ values in \textcolor{black}{Fig. \ref{fig:Complexity_BS_UE}(b)}, the proposed {ELM-based} approach using $(\bdzeta_\rmBS^{\rmS_2\rmI_2},\bdgamma_\rmBS^{\rmS_2})$ provides lower {computational} complexity than that with $(\bdzeta_\rmBS^{\rmS_2\rmI_1},\bdgamma_\rmBS^{\rmS_2})$.
By taking another look at \textcolor{black}{Fig. \ref{fig:NMSE_M_BS}(a)}, one can notice that the corresponding estimation performance of the proposed RE-ELM with $(\bdzeta_\rmBS^{\rmS_2\rmI_2},\bdgamma_\rmBS^{\rmS_2})$ {outperforms} that with $(\bdzeta_\rmBS^{\rmS_2\rmI_1},\bdgamma_\rmBS^{\rmS_2})$.
Thus, the second type of input-output pair is preferable in such a case.
Except for estimating $\bfB_k$ with large $M$ values, the proposed ELM-based approach adopting the first type of input-output pair is recommended due to its lower cost.

\section{Conclusion}\label{sec:conclusion}
In this paper, an ELM-based two-stage channel estimation approach has been proposed for an IRS-assisted multi-user ISAC system.
The proposed {ELM-based} framework, which involves the DE-ELM and RE-ELM structures, has been employed to estimate the direct and reflected SAC channels successively.
The pilot transmission policy of the FD ISAC BS, multiple uplink UEs, and IRS has been designed to support SAC channels estimation.
Two types of input-output pairs have been devised for the ELMs.
Numerical results have shown that under various SNR setups, the proposed {ELM-based} approach exhibits excellent generalization capacity and attains significant enhancement in estimation accuracy over the LS benchmark scheme; \textcolor{black}{{it} achieves more than $10^{1.5}\,\rm x$ and $10^{1.4}\,\rm x$ NMSE improvements at $\text{SNR}=-10\,\rm dB$, respectively, for the channel estimation at the ISAC BS and downlink UEs.}
\textcolor{black}{Compared to the \textcolor{black}{NN}-based benchmark schemes, the estimation accuracy of the proposed ELM-based approach outperforms the CNN-based benchmark, while it is comparable to the FNN-based one.
The corresponding learning speed of the proposed ELM-based approach is extremely fast (i.e., less than $15$ seconds) and outperforms both the CNN-based and FNN-based benchmark schemes.
}
Moreover, under different channel dimensions, the proposed {ELM-based} approach provides a superior estimation accuracy and comparable computational complexity compared to the LS benchmark scheme.
This indicates that the proposed {ELM-based} approach {can be used for} channel {estimation} {at} both ISAC BS and downlink UEs to meet their high estimation accuracy and low-cost requirements.
As potential directions for further investigation, the proposed ELM-based approach will be extended to more practical scenarios, \textcolor{black}{which consider} the imperfect synchronization between the SAC signals and \textcolor{black}{the} residual SI, \textcolor{black}{the} non-linear residual SI model, \textcolor{black}{time-varying SAC channels}, as well as wideband systems.



\bibliographystyle{IEEEtran}

\bibliography{Reference_ISAC}

\begin{thebibliography}{10}
\providecommand{\url}[1]{#1}
\csname url@samestyle\endcsname
\providecommand{\newblock}{\relax}
\providecommand{\bibinfo}[2]{#2}
\providecommand{\BIBentrySTDinterwordspacing}{\spaceskip=0pt\relax}
\providecommand{\BIBentryALTinterwordstretchfactor}{4}
\providecommand{\BIBentryALTinterwordspacing}{\spaceskip=\fontdimen2\font plus
\BIBentryALTinterwordstretchfactor\fontdimen3\font minus
  \fontdimen4\font\relax}
\providecommand{\BIBforeignlanguage}[2]{{%
\expandafter\ifx\csname l@#1\endcsname\relax
\typeout{** WARNING: IEEEtran.bst: No hyphenation pattern has been}%
\typeout{** loaded for the language `#1'. Using the pattern for}%
\typeout{** the default language instead.}%
\else
\language=\csname l@#1\endcsname
\fi
#2}}
\providecommand{\BIBdecl}{\relax}
\BIBdecl

\bibitem{ref-new-IRS-defi}
M.~A. ElMossallamy, H.~Zhang, L.~Song, K.~G. Seddik, Z.~Han, and G.~Y. Li,
  ``Reconfigurable intelligent surfaces for wireless communications:
  Principles, challenges, and opportunities,'' \emph{IEEE Trans. Cogn. Commun.
  and Netw.}, vol.~6, no.~3, pp. 990--1002, Sep. 2020.

\bibitem{ref:IRS-SCMA-optimize}
I.~Al-Nahhal, O.~A. Dobre, E.~Basar, T.~M.~N. Ngatched, and S.~Ikki,
  ``Reconfigurable intelligent surface optimization for uplink sparse code
  multiple access,'' \emph{IEEE Commun. Lett.}, vol.~26, no.~1, pp. 133--137,
  Jan. 2022.

\bibitem{ref:IRS-SCMA}
I.~Al-Nahhal, O.~A. Dobre, and E.~Basar, ``Reconfigurable intelligent
  surface-assisted uplink sparse code multiple access,'' \emph{IEEE Commun.
  Lett.}, vol.~25, no.~6, pp. 2058--2062, Jun. 2021.

\bibitem{ref:ChModel-refpower}
A.~Faisal, I.~Al-Nahhal, O.~A. Dobre, and T.~M.~N. Ngatched, ``Deep
  reinforcement learning for optimizing {RIS}-assisted {HD-FD} wireless
  systems,'' \emph{IEEE Commun. Lett.}, vol.~25, no.~12, pp. 3893--3897, Dec.
  2021.

\bibitem{ref:new-IRS-passive-beam}
M.~Yue, L.~Liu, and X.~Yuan, ``Practical {RIS}-aided coded systems: Joint
  precoding and passive beamforming,'' \emph{IEEE Wireless Commun. Lett.},
  vol.~10, no.~11, pp. 2345--2349, Nov. 2021.

\bibitem{ref:IRS-beam-gain3}
P.~Wang, J.~Fang, X.~Yuan, Z.~Chen, and H.~Li, ``Intelligent reflecting
  surface-assisted millimeter wave communications: Joint active and passive
  precoding design,'' \emph{IEEE Trans. Veh. Technol.}, vol.~69, no.~12, pp.
  14\,960--14\,973, Dec. 2020.

\bibitem{ref:IRS-turn-off}
X.~Wei, D.~Shen, and L.~Dai, ``Channel estimation for {RIS} assisted wireless
  communications--{Part I}: Fundamentals, solutions, and future
  opportunities,'' \emph{IEEE Commun. Lett.}, vol.~25, no.~5, pp. 1398--1402,
  May 2021.

\bibitem{ref:new-IRS-Challenge}
W.~Zhang, J.~Xu, W.~Xu, D.~W.~K. Ng, and H.~Sun, ``Cascaded channel estimation
  for {IRS}-assisted mm{W}ave multi-antenna with quantized beamforming,''
  \emph{IEEE Commun. Lett.}, vol.~25, no.~2, pp. 593--597, Feb. 2021.

\bibitem{ref:IRS-ChE-onoff1}
D.~Mishra and H.~Johansson, ``Channel estimation and low-complexity beamforming
  design for passive intelligent surface assisted {MISO} wireless energy
  transfer,'' in \emph{Proc. IEEE ICASSP}, May 2019, pp. 4659--4663.

\bibitem{ref:new-IRS-ChE-DFT}
T.~L. Jensen and E.~D. Carvalho, ``An optimal channel estimation scheme for
  intelligent reflecting surfaces based on a minimum variance unbiased
  estimator,'' in \emph{Proc. IEEE ICASSP}, May 2020, pp. 5000--5004.

\bibitem{ref:IRS-ChE-elegroup2}
S.~Zhang and R.~Zhang, ``Capacity characterization for intelligent reflecting
  surface aided {MIMO} communication,'' \emph{IEEE J. Sel. Areas Commun.},
  vol.~38, no.~8, pp. 1823--1838, Aug. 2020.

\bibitem{ref:DL-IRS-ChE-WCL}
A.~M. Elbir., A.~Papazafeiropoulos, P.~Kourtessis, and S.~Chatzinotas, ``Deep
  channel learning for large intelligent surfaces aided mm-wave massive {MIMO}
  systems,'' \emph{IEEE Wireless Commun. Lett.}, vol.~9, no.~9, pp. 1447--1451,
  May 2020.

\bibitem{ref:IRS-ChE-CNN-group}
S.~Zhang, S.~Zhang, F.~Gao, J.~Ma, and O.~A. Dobre, ``Deep learning-based {RIS}
  channel extrapolation with element-grouping,'' \emph{IEEE Wireless Commun.
  Lett.}, vol.~10, no.~12, pp. 2644--2648, Dec. 2021.

\bibitem{ref:ELM-OFDM-CE}
C.~Qing, L.~Wang, L.~Dong, and J.~Wang, ``Enhanced {ELM} based channel
  estimation for {RIS}-assisted {OFDM} systems with insufficient cp and
  imperfect hardware,'' \emph{IEEE Commun. Lett.}, vol.~26, no.~1, pp.
  153--157, Jan. 2022.

\bibitem{ref:DL-IRS-ChE-TWC}
C.~Liu, X.~Liu, D.~W.~K. Ng, and J.~Yuan, ``Deep residual learning for channel
  estimation in intelligent reflecting surface-assisted multi-user
  communications,'' \emph{IEEE Trans. Wireless Commun.}, vol.~21, no.~2, pp.
  898--912, Feb. 2022.

\bibitem{ref:ISAC-survey}
F.~Liu, Y.~Cui, C.~Masouros, J.~Xu, T.~X. Han, Y.~C. Eldar, and S.~Buzzi,
  ``Integrated sensing and communications: Toward dual-functional wireless
  networks for 6{G} and beyond,'' \emph{IEEE J. Sel. Areas Commun.}, vol.~40,
  no.~6, pp. 1728--1767, Jun. 2022.

\bibitem{ref:ChModel-b}
F.~Liu, C.~Masouros, A.~P. Petropulu, H.~Griffiths, and L.~Hanzo, ``Joint radar
  and communication design: Applications, state-of-the-art, and the road
  ahead,'' \emph{IEEE Trans. Commun.}, vol.~68, no.~6, pp. 3834--3862, Jun.
  2020.

\bibitem{ref:ISAC-improve-S1}
B.~K. Chalise, M.~G. Amin, and B.~Himed, ``Performance tradeoff in a unified
  passive radar and communications system,'' \emph{IEEE Signal Process. Lett.},
  vol.~24, no.~9, pp. 1275--1279, Sep. 2017.

\bibitem{ref:ISAC-improve-S2}
F.~Liu, Y.~Liu, A.~Li, C.~Masouros, and Y.~C. Eldar, ``Cramer-rao bound
  optimization for joint radar-communication beamforming,'' \emph{IEEE Trans.
  Signal Process.}, vol.~70, pp. 240--253, 2022.

\bibitem{ref:new-ISAC-improve-S2}
X.~Liu, T.~Huang, N.~Shlezinger, Y.~Liu, J.~Zhou, and Y.~C. Eldar, ``Joint
  transmit beamforming for multiuser {MIMO} communications and {MIMO} radar,''
  \emph{IEEE Trans. Signal Process.}, vol.~68, pp. 3929--3944, 2020.

\bibitem{ref:ISAC-improve-C1}
F.~Liu, W.~Yuan, C.~Masouros, and J.~Yuan, ``Radar-assisted predictive
  beamforming for vehicular links: Communication served by sensing,''
  \emph{IEEE Trans. Wireless Commun.}, vol.~19, no.~11, pp. 7704--7719, Nov.
  2020.

\bibitem{ref:ISAC-improve-C2}
W.~Yuan, F.~Liu, C.~Masouros, J.~Yuan, D.~W.~K. Ng, and N.~González-Prelcic,
  ``Bayesian predictive beamforming for vehicular networks: A low-overhead
  joint radar-communication approach,'' \emph{IEEE Trans. Wireless Commun.},
  vol.~20, no.~3, pp. 1442--1456, Mar. 2021.

\bibitem{ref:new-ISAC-improve-C3}
W.~Shi, W.~Xu, X.~You, C.~Zhao, and K.~Wei, ``Intelligent reflection enabling
  technologies for integrated and green internet-of-everything beyond 5g:
  Communication, sensing, and security,'' \emph{IEEE Wireless Commun.}, Early
  access, 2022.

\bibitem{ref:ISAC-ChEst-RSU}
A.~Ali, N.~Gonzalez-Prelcic, R.~W. Heath, and A.~Ghosh, ``Leveraging sensing at
  the infrastructure for mm{W}ave communication,'' \emph{IEEE Commun. Mag.},
  vol.~58, no.~7, pp. 84--89, Jul. 2020.

\bibitem{ref:ChModel-pathloss-SJ}
Z.~Jiang, M.~Rihan, P.~Zhang, L.~Huang, Q.~Deng, J.~Zhang, and E.~M. Mohamed,
  ``Intelligent reflecting surface aided dual-function radar and communication
  system,'' \emph{IEEE Syst. J.}, vol.~16, no.~1, pp. 475--486, Mar. 2022.

\bibitem{ref:Joint-ISAC-IRS-beam}
F.~Wang, H.~Li, and J.~Fang, ``Joint active and passive beamforming for
  {IRS}-assisted radar,'' \emph{IEEE Signal Process. Lett.}, vol.~29, pp.
  349--353, 2022.

\bibitem{ref:new-Joint-ISAC-IRS-beam1}
X.~Song, D.~Zhao, H.~Hua, T.~X. Han, X.~Yang, and J.~Xu, ``Joint transmit and
  reflective beamforming for {IRS}-assisted integrated sensing and
  communication,'' in \emph{Proc. IEEE WCNC}, May 2022, pp. 189--194.

\bibitem{ref:IRS-ISAC-wavebeam1}
X.~Wang, Z.~Fei, Z.~Zheng, and J.~Guo, ``Joint waveform design and passive
  beamforming for {RIS}-assisted dual-functional radar-communication system,''
  \emph{IEEE Trans. Veh. Technol.}, vol.~70, no.~5, pp. 5131--5136, May 2021.

\bibitem{ref:IRS-ISAC-wavebeam2}
X.~Wang, Z.~Fei, J.~Huang, and H.~Yu, ``Joint waveform and discrete phase shift
  design for {RIS}-assisted integrated sensing and communication system under
  cramer-rao bound constraint,'' \emph{IEEE Trans. Veh. Technol.}, vol.~71,
  no.~1, pp. 1004--1009, Jan. 2022.

\bibitem{ref:IRS-ISAC-ChEst-TVT}
Y.~Liu, I.~Al-Nahhal, O.~A. Dobre, and F.~Wang, ``Deep-learning channel
  estimation for {IRS}-assisted integrated sensing and communication system,''
  \emph{submitted to IEEE Trans. Veh. Technol.}, 2022.

\bibitem{ref:ELM1}
G.~Huang, Q.~Zhu, and C.~Siew, ``Extreme learning machine: theory and
  applications,'' \emph{Neurocomputing}, vol.~70, no. 1-3, pp. 489--501, May
  2006\color{black}.

\bibitem{ref:EM-application1}
\color{black}S. Ding, X.~Xu, and R.~Nie, ``Extreme learning machine and its
  applications,'' \emph{Neural. Comput. Appl.}, vol.~25, no.~3, pp. 549--556,
  Sep. 2014.

\bibitem{ref:EM-application2}
\color{black}G. Song, Q.~Dai, X.~Han, and L.~Guo, ``Two novel {ELM}-based
  stacking deep models focused on image recognition,'' \emph{Appl. Intell.},
  vol.~50, no.~5, pp. 1345--1366, May 2020.

\bibitem{ref:major-SI-estimation3}
\color{black}M. Elsayed, A.~A.~A. El-Banna, O.~A. Dobre, W.~Shiu, and P.~Wang,
  ``Hybrid-layers neural network architectures for modeling the
  self-interference in full-duplex systems,'' \emph{IEEE Trans. Vehicular
  Technol.}, vol.~71, no.~6, pp. 6291--6307, Jun. 2022.

\bibitem{ref:major-SI-estimation4}
M.~Elsayed, A.~A.~A. El-Banna, O.~A. Dobre, W.~Shiu, and P.~Wang, ``Full-duplex
  self-interference cancellation using dual-neurons neural networks,''
  \emph{IEEE Commun. Lett.}, vol.~26, no.~3, pp. 557--561, Mar. 2022.

\bibitem{ref:major-SI-estimation5}
------, ``Low complexity neural network structures for self-interference
  cancellation in full-duplex radio,'' \emph{IEEE Commun. Lett.}, vol.~25,
  no.~1, pp. 181--185, Jan. 2021.

\bibitem{ref:major-3gpp.36.211}
3GPP, ``{Evolved Universal Terrestrial Radio Access (E-UTRA); Physical channels
  and modulation},'' {3GPP}, TS 36.211, Jun. 2016.

\bibitem{ref:major-3.5G}
Z.~Xiao and Y.~Zeng, ``Waveform design and performance analysis for full-duplex
  integrated sensing and communication,'' \emph{IEEE J. Sel. Areas Commun.},
  vol.~40, no.~6, pp. 1823--1837, Jun. 2022\color{black}.

\bibitem{ref:Bayesian-pilot-num}
J.~Ma, S.~Zhang, H.~Li, F.~Gao, and S.~Jin, ``Sparse bayesian learning for the
  time-varying massive {MIMO} channels: Acquisition and tracking,'' \emph{IEEE
  Trans. Commun.}, vol.~67, no.~3, pp. 1925--1938, Mar. 2019.

\bibitem{ref:data-augment}
C.~Shorten and T.~M. Khoshgoftaar, ``A survey on image data augmentation for
  deep learning,'' \emph{J. Big Data}, vol.~6, no.~1, pp. 1--48, Jul. 2019.

\bibitem{ref:matrix-inverse}
R.~W. Farebrother, \emph{Linear Least Squares Computations}.\hskip 1em plus
  0.5em minus 0.4em\relax Routledge, 1988\color{black}.

\bibitem{ref:major-coherent-time}
S.~Biswas, C.~Masouros, and T.~Ratnarajah, ``Performance analysis of large
  multiuser {MIMO} systems with space-constrained 2-{D} antenna arrays,''
  \emph{IEEE Trans. Wireless Commun.}, vol.~15, no.~5, pp. 3492--3505, May
  2016\color{black}.

\bibitem{ref:channel-b-fixed-amplitude}
N.~Nartasilpa, A.~Salim, D.~Tuninetti, and N.~Devroye, ``Communications system
  performance and design in the presence of radar interference,'' \emph{IEEE
  Trans. Commun.}, vol.~66, no.~9, pp. 4170--4185, Sep. 2018.

\bibitem{ref:ChModel-power}
T.~Jiang, H.~V. Cheng, and W.~Yu, ``Learning to reflect and to beamform for
  intelligent reflecting surface with implicit channel estimation,'' \emph{IEEE
  J. Sel. Areas Commun.}, vol.~39, no.~7, pp. 1931--1945, Jul. 2021.

\end{thebibliography}

\begin{IEEEbiography}[{\includegraphics[width=1in,height=1.25in,clip,keepaspectratio]{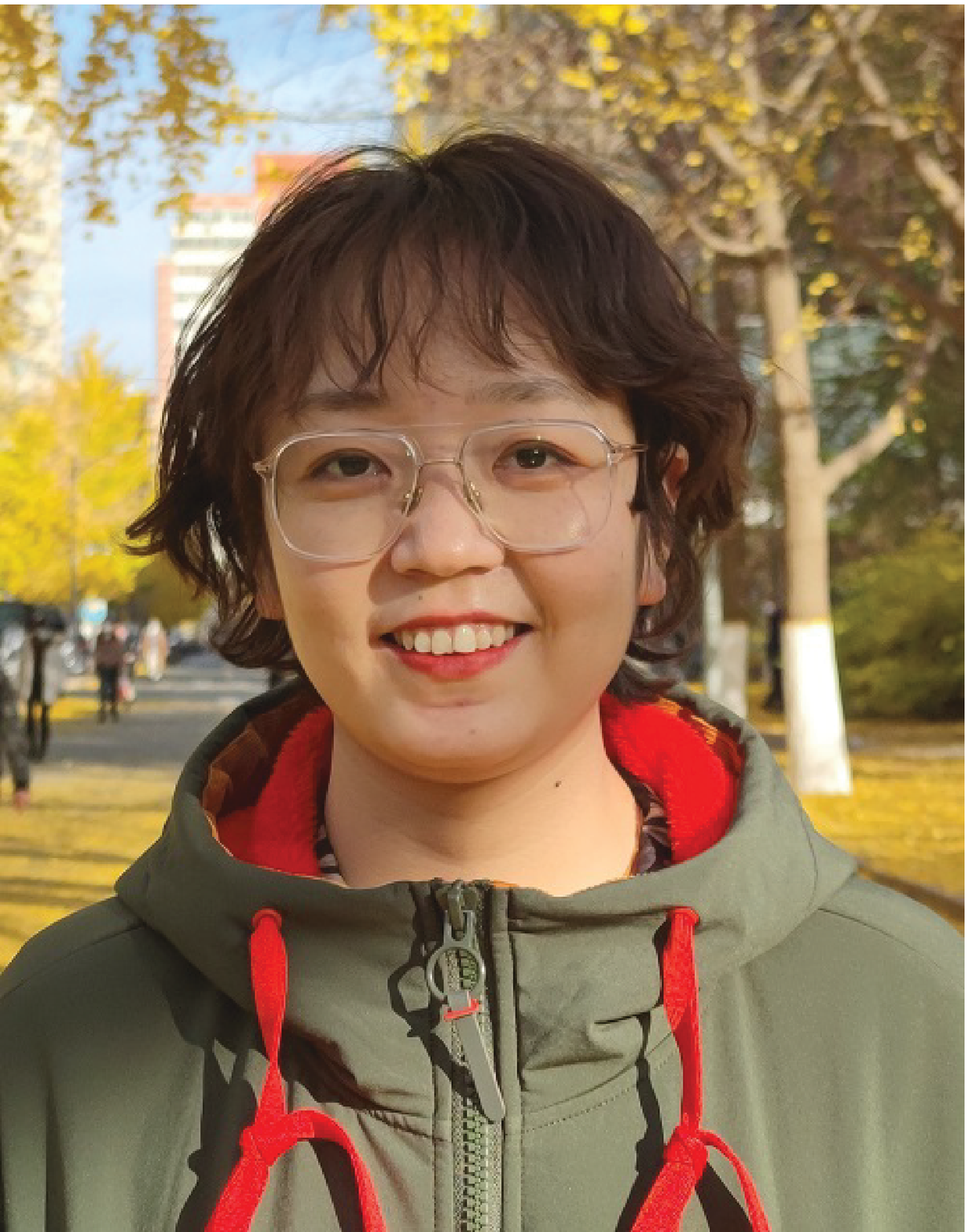}}]{Yu Liu} received the B.Eng. degree from the School of Electronic and Information Engineering, Beijing Jiaotong University, Beijing, China, in 2017. She is currently pursuing the Ph.D. degree with the State Key Laboratory of Rail Traffic Control and Safety, Beijing Jiaotong University, Beijing, China. Her current research interests include signal processing, interference cancellation, and integrated sensing and communication.
\end{IEEEbiography}

\begin{IEEEbiography}[{\includegraphics[width=1in,height=1.25in,clip,keepaspectratio]{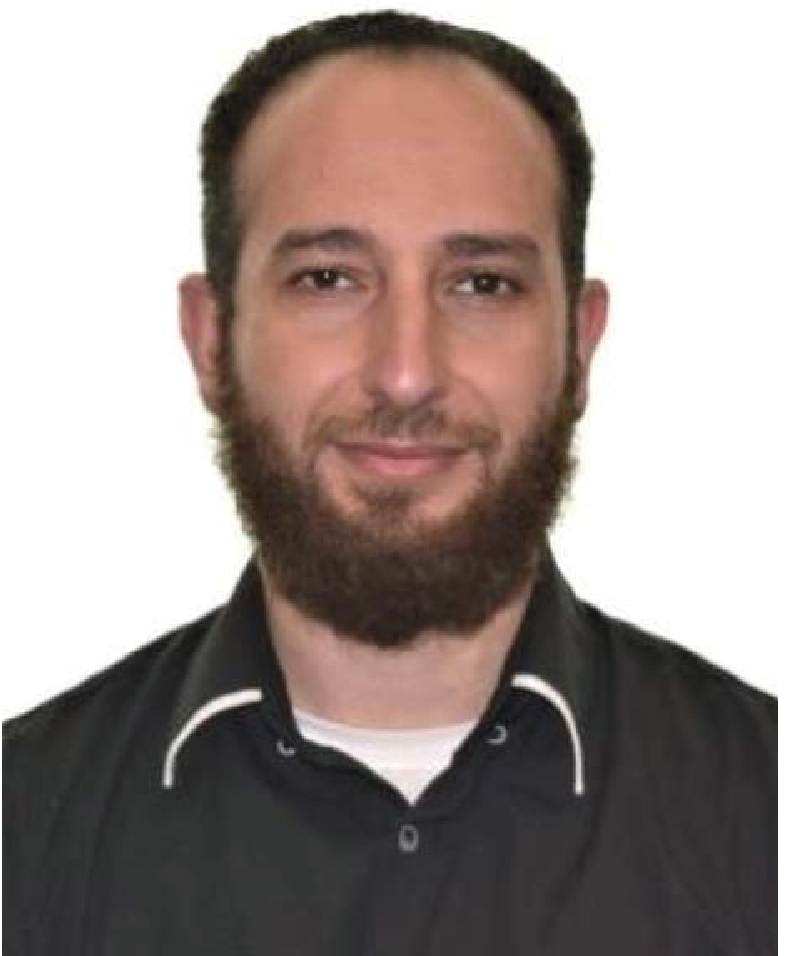}}]{Ibrahim Al-Nahhal}
(IEEE senior member) is a research associate and per-course instructor at Memorial University, Canada, since 2021. He received the B.Sc. (Honours), M.Sc., and Ph.D. degrees in Electronics and Communications Engineering from Al-Azhar University in Cairo, Egypt-Japan University for Science and Technology, Egypt, Memorial University, Canada, in 2007, 2014, and 2020, respectively. Between 2008 and 2012, he was an engineer in industry, and a Teaching Assistant at the Faculty of Engineering, Al-Azhar University in Cairo, Egypt. From 2014 to 2015, he was a physical layer expert at Nokia (formerly Alcatel-Lucent), Belgium. He holds three patents. He co-authored 30+ peer-reviewed journals and conference papers in top-ranked venues. He serves as Editor of IEEE Wireless Communications Letters. He served as a Technical Program Committee and Reviewer for various prestigious journals and conferences. He was awarded the Exemplary Reviewer of IEEE Communications Letters in 2017. His research interests are reconfigurable intelligent surfaces, full-duplex communications, integrated sensing and communication, channel estimation, machine learning, design of low-complexity receivers for emerging technologies, spatial modulation, multiple-input multiple-output communications, sparse code multiple access, and optical communications.
\end{IEEEbiography}

\begin{IEEEbiography}[{\includegraphics[width=1in,height=1.25in,clip,keepaspectratio]{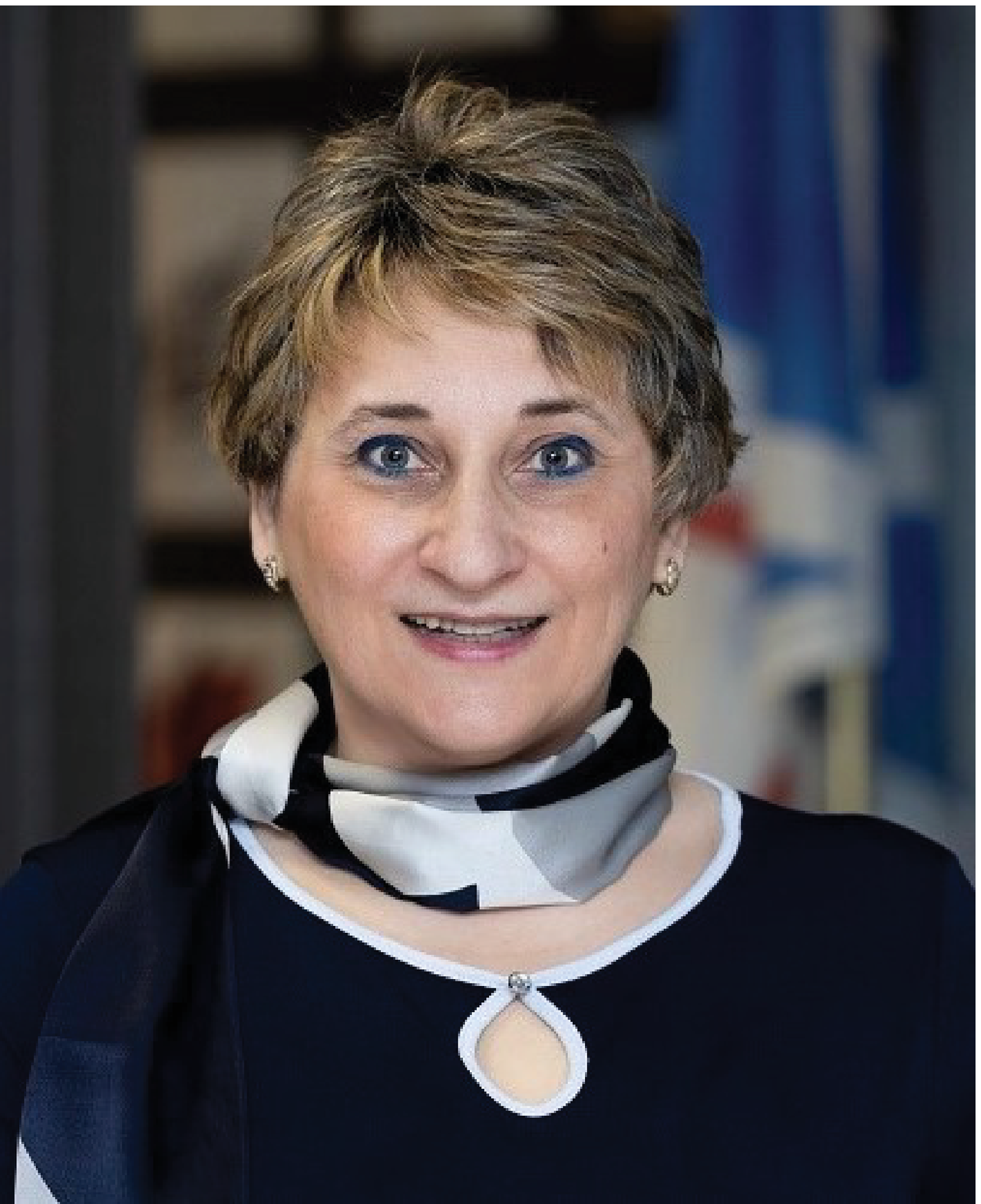}}]{Octavia A. Dobre} (Fellow, IEEE) received the Dipl. Ing. and Ph.D. degrees from the Polytechnic Institute of Bucharest, Romania, in 1991 and 2000, respectively. Between 2002 and 2005, she was with the New Jersey Institute of Technology, USA. In 2005, she joined Memorial University, Canada, where she is currently a Professor and Canada Research Chair Tier 1. She was a Visiting Professor with Massachusetts Institute of Technology, USA and Universit{$\acute {\text e}$} de Bretagne Occidentale, France.
Her research interests encompass wireless communication and networking technologies, as well as optical and underwater communications. She has (co-)authored over 450 refereed papers in these areas.
Dr. Dobre serves as the Director of Journals of the Communications Society. She was the inaugural Editor-in-Chief (EiC) of the IEEE Open Journal of the Communications Society and the EiC of the IEEE Communications Letters.
Dr. Dobre was a Fulbright Scholar, Royal Society Scholar, and Distinguished Lecturer of the IEEE Communications Society. She obtained Best Paper Awards at various conferences, including IEEE ICC, IEEE Globecom, IEEE WCNC, and IEEE PIMRC. Dr. Dobre is an elected member of the European Academy of Sciences and Arts, a Fellow of the Engineering Institute of Canada, and a Fellow of the Canadian Academy of Engineering.
\end{IEEEbiography}

\begin{IEEEbiography}[{\includegraphics[width=1in,height=1.25in,clip,keepaspectratio]{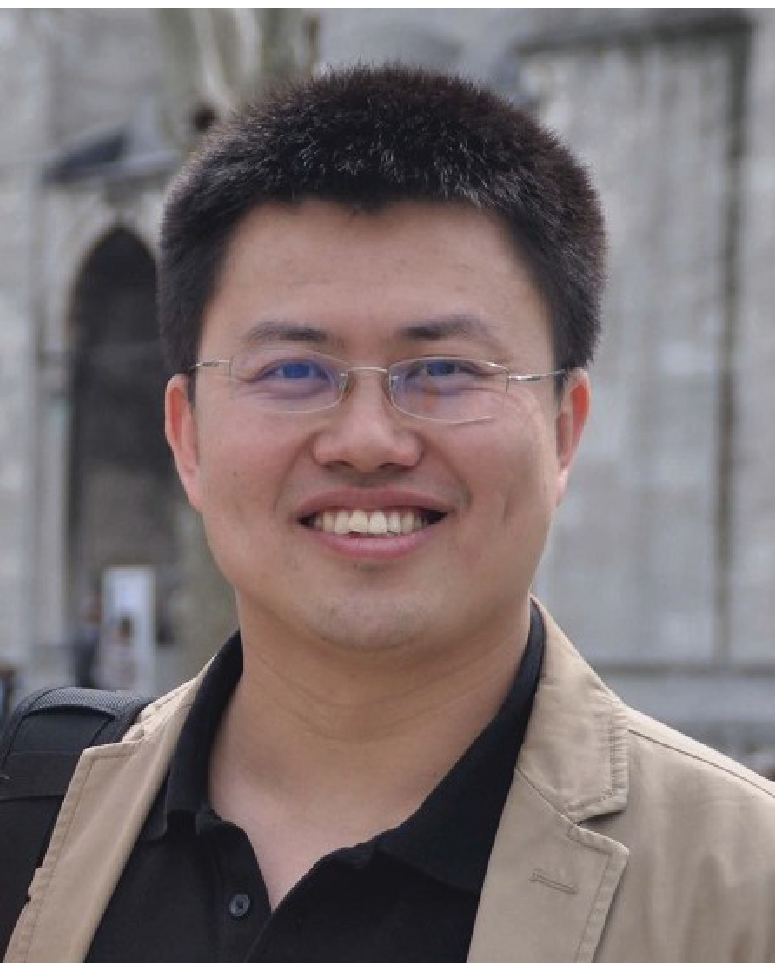}}]{Fanggang Wang} (S'10-M'11-SM'16) received the B.Eng. and Ph.D. degrees from the School of Information and Communication Engineering, Beijing University of Posts and Telecommunications, Beijing, China, in 2005 and 2010, respectively. He was a Post-Doctoral Fellow with the Institute of Network Coding, The Chinese University of Hong Kong, Hong Kong, from 2010 to 2012. He was a Visiting Scholar with the Massachusetts Institute of Technology from 2015 to 2016 and the Singapore University of Technology and Design in 2014. He is currently a Professor with the State Key Laboratory of Rail Traffic Control and Safety, School of Electronic and Information Engineering, Beijing Jiaotong University. His research interests are in wireless communications, signal processing, and information theory. He served as an Editor for the IEEE COMMUNICATIONS LETTERS and technical program committee member for several conferences.
\end{IEEEbiography}

\begin{IEEEbiography}[{\includegraphics[width=1in,height=1.25in,clip,keepaspectratio]{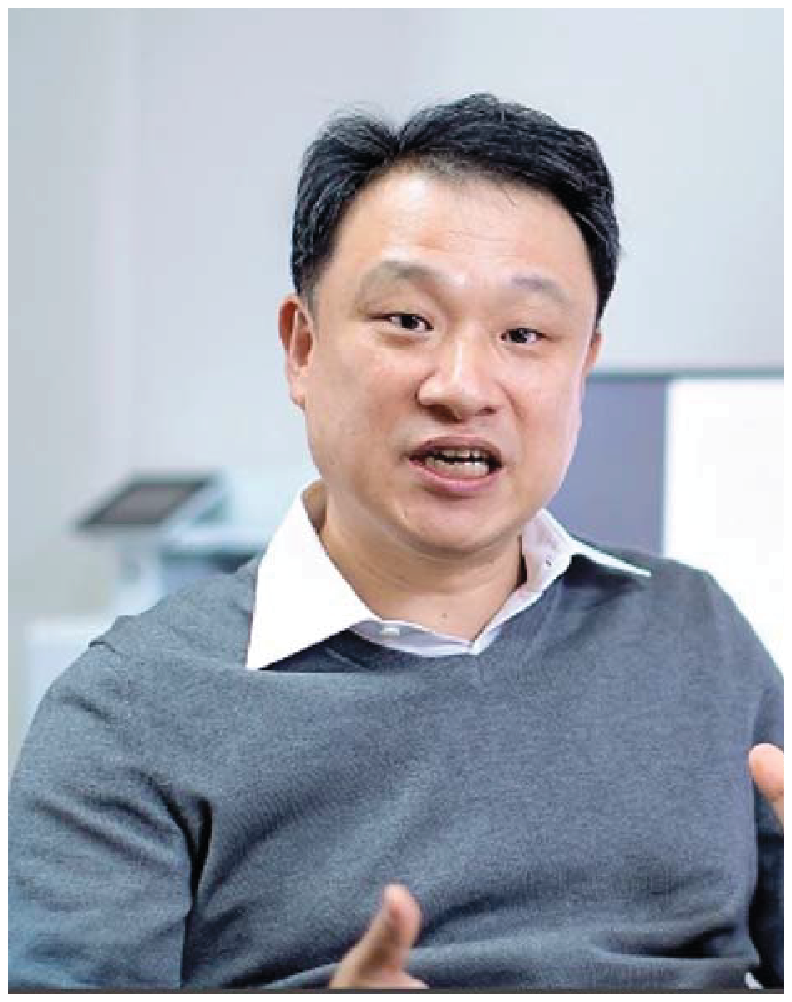}}]{Hyundong Shin} (Fellow, IEEE) received the B.S. degree in Electronics Engineering from Kyung Hee University (KHU), Yongin-si, Korea, in 1999, and the M.S. and Ph.D. degrees in Electrical Engineering from Seoul National University, Seoul, Korea, in 2001 and 2004, respectively. During his postdoctoral research at the Massachusetts Institute of Technology (MIT) from 2004 to 2006, he was with the Laboratory for Information Decision Systems (LIDS). In 2006, he joined the KHU, where he is currently a Professor in the Department of Electronic Engineering. His research interests include quantum information science, wireless communication, and machine intelligence.
Dr. Shin received the IEEE Communications Society's Guglielmo Marconi Prize Paper Award and William R. Bennett Prize Paper Award. He served as the Publicity Co-Chair for the IEEE PIMRC and the Technical Program Co-Chair for the IEEE WCNC and the IEEE GLOBECOM. He was an Editor of IEEE Transactions on Wireless Communications and IEEE Communications Letters.
\end{IEEEbiography}

\end{document}